\documentclass[leqno,draft,11pt,a4paper]{article}
\usepackage{amssymb,amsmath,latexsym,theorem}
\usepackage[includeheadfoot,margin=1in]{geometry}

\allowdisplaybreaks[2]

\newcommand\LOR{\bigvee}
\newcommand\ET{\bigwedge}
\newcommand\model{\vDash}
\newcommand\nmodel{\nvDash}

\newcommand\fii{\varphi}
\newcommand\roo{\varrho}
\newcommand\ep{\varepsilon}

\newcommand\p[1]{\langle#1\rangle}
\newcommand\delim[4]{\ifx X#3X\left#1#4\right#2\else\csname#3l\endcsname#1#4\csname#3r\endcsname#2\fi}
\newcommand\pp[2][]{\delim<>{#1}{#2}}
\newcommand\lh[1]{\lvert#1\rvert}
\newcommand\LH[2][]{\delim||{#1}{#2}}
\newcommand\dlh[1]{\lVert#1\rVert}
\newcommand\Dlh[2][]{\delim\lVert\rVert{#1}{#2}}
\newcommand\bez{\smallsetminus}
\newcommand\sset{\subseteq}
\newcommand\ssset{\subsetneq}
\newcommand\pw[1]{\mathcal P(#1)}
\newcommand\nul{\varnothing}
\newcommand\res{\mathbin\restriction}
\newcommand\two{\mathbf2}

\newcommand\cl[1]{\lceil#1\rceil}

\newcommand\ru{\mathrel/}
\newcommand\sls{|^{}}
\newcommand\Sls{\|^{}}
\newcommand\nsls{\mathord\nmid^{}}
\newcommand\nSls{\mathord\nparallel^{}}

\DeclareMathOperator\dom{dom}
\DeclareMathOperator\cls{cl}

\newcommand\ccp{\mathrm{CC}}
\newcommand\itp{\mathrm{Itp}}

\newcommand\lgc[1]{\mathbf{#1}}
\newcommand\IPC{\lgc{IPC}}
\newcommand\Form{\mathrm{Form}}
\newcommand\Var{\mathrm{Var}}
\newcommand\nmi{\mathrm{NM}_\to}
\newcommand\fri{\mathrm{F}_\to}
\newcommand\imi{\mathrm{({\to}I)}}
\newcommand\ime{\mathrm{({\to}E)}}

\newcommand\ob[1]{\overline{#1}}
\newcommand\mclap[1]{\mathpalette\doclap{#1}}
\newcommand\doclap[2]{\hbox to0pt{\hss$#1#2$\hss}}
\newcommand\bret[1]{\land^{\mclap{#1}}_{\vphantom l}}
\newcommand\RET[2]{\mathop{\ET\nolimits^{\mclap{#1}}}\displaylimits_{#2}}

\def\bme{\hskip.75em\relax}
\def\noproof{\relax\ifmmode\eqno\Box\else\leavevmode\unskip\bme\vadjust{}\nobreak\hfill$\Box$\par\fi}
\newenvironment{Pf}[1][]
  {\par\noindent\textit{Proof:}\bme\ignorespaces}
  {\noproof\pagebreak[2]\vskip\medskipamount\ignorespacesafterend}

\theoremstyle{plain}
\newtheorem{Thm}{Theorem}[section]
\newtheorem{Lem}[Thm]{Lemma}
\newtheorem{Obs}[Thm]{Observation}
\theorembodyfont\upshape
\newtheorem{Def}[Thm]{Definition}
\newtheorem{Rem}[Thm]{Remark}

\begingroup \lccode`\~=`\/ \lowercase{ \endgroup
  \providecommand\url{\begingroup \catcode`\~=12 \catcode`\_=12 \catcode`\/=13 \let~\beginslash \finishurl}
  \newcommand\finishurl[1]{\texttt{#1}\endgroup}
  \newcommand\beginslash{/\futurelet\nexttoken\finishslash}
  \newcommand\finishslash{\ifx\nexttoken~\else\penalty\relpenalty\fi}
}

\author{Emil Je\v r\'abek\\[\medskipamount]
Institute of Mathematics, Czech Academy of Sciences\\
\small \v Zitn\'a 25,
115\:67 Praha 1,
Czech Republic,
email: \texttt{jerabek@math.cas.cz}
}

\title{A simplified lower bound for implicational logic}

\begin{document}
\maketitle

\begin{abstract}
We present a streamlined and simplified exponential lower bound on the length of proofs in intuitionistic implicational
logic, adapted to Gordeev and Haeusler's dag-like natural deduction.
\end{abstract}

\section{Introduction}\label{sec:introduction}

Frege proof systems (often called Hilbert-style systems outside proof complexity) are among the simplest and most
natural proof systems for classical and nonclassical propositional logics. By results of Reckhow and Cook
\cite{reck,cookrek}, all classical Frege systems are not only polynomially equivalent to each other, but also to natural
deduction systems and to sequent calculi (with cut), which is further testimony to their robustness and fundamental
status. Although it is commonly assumed for all classical propositional proof systems that some tautologies require
exponentially large proofs, this has been proven so far only for relatively weak proof systems, such as constant-depth
Frege, polynomial calculus, and cutting planes (see e.g.\ \cite{kra:pfcomp,buss-nord:pfcomp}). Unrestricted Frege
systems are far beyond the reach of current techniques: nothing better is known than a linear lower bound on the number
of proof lines and a quadratic bound on the overall proof size \cite{buss:rems,book}.

Interestingly, the state of affairs is much better in nonclassical logics:
Hrube\v{s}~\cite{hru:lbmod,hru:lbint,hru:nonclas} proved exponential lower bounds on the number of lines in Frege
proofs for some modal logics and intuitionistic logic, which was generalized by Je\v r\'abek~\cite{ej:sfef} to all
transitive modal and superintuitionistic logics with unbounded branching, and by Jalali~\cite{jal:substr} to
substructural logics. Even though the techniques are based on variants of the feasible disjunction property (i.e.,
given a proof of $\fii\lor\psi$, we can decide in polynomial time which of $\fii$ or~$\psi$ is provable), and as such
ostensibly require disjunction, Je\v r\'abek~\cite{ej:implic} showed that the superintuitionistic exponential lower
bounds hold for a sequence of purely implicational intuitionistic tautologies.

In a series of papers, Gordeev and Haeusler~\cite{gor-hae:crap-i,gor-hae:crap-ii,gor-hae:crap-ii+,gor-hae:crap-pfth}
claim to prove that all intuitionistic implicational tautologies have polynomial-size proofs in a dag-like version of
(Gentzen/Prawitz-style) natural deduction, which---if true---would imply $\mathrm{NP}=\mathrm{PSPACE}$. These claims are wrong, as they
contradict the above-mentioned exponential lower bounds on the length of proofs of implicational tautologies in
intuitionistic proof systems. Unfortunately, this fact may not be so obvious to readers unfamiliar with nonclassical
proof complexity literature, and in any event, the full proof of the lower bound requires tracking down multiple
papers: the Frege lower bound for implicational tautologies in~\cite{ej:implic} builds on a lower bound for
unrestricted intuitionistic tautologies, as proved in either of~\cite{hru:lbint,hru:nonclas,ej:sfef}; these in turn
rely on an exponential lower bound on the size of monotone circuits separating the Clique--Colouring disjoint NP
pair which---in view of an observation of Tardos~\cite{tard-mon}---follows from Alon and Boppana~\cite{alonbop}
(improving a superpolynomial lower bound by Razborov~\cite{razb:monot}). Finally, one needs a polynomial simulation of
natural deduction by Frege systems: this is originally due to Reckhow and Cook~\cite{reck,cookrek}, but they state it
for a sequent-style formulation of natural deduction rather than Prawitz-style, let alone the further variant
introduced only recently by Gordeev and Haeusler; while it is clear to a proof complexity practitioner that the
argument can be easily adapted to all such variants, this is, strictly speaking, not explicitly proved in any extant
literature.

The primary goal of this paper is to give a simple direct proof of an exponential lower bound on the length of proofs
of intuitionistic implicational tautologies in Gordeev and Haeusler's dag-like natural deduction. The streamlined
argument replaces all proof-theoretic components of the lower bound mentioned above (intuitionistic lower bound,
reduction to implicational logic, simulation of natural deduction by Frege), thus it is self-contained except for the
combinatorial component (i.e., a monotone circuit lower bound; to simplify our tautologies, we will use a lower bound
by Hrube\v s and Pudl\'ak~\cite{hru-pud:col-col} instead of Alon--Boppana). It is based on the efficient Kleene
slash approach employed in \cite{fff,min-koj,ej:modfrege,ej:sfef}. While we strive to keep the proof of the main result
as simple as possible, we also briefly indicate how to generalize it to recover almost the full strength of the
lower bound from~\cite{ej:implic}.

The intended audience of the paper is twofold:
\begin{itemize}
\item Readers with some general background in logic or computer science, but unfamiliar with proof complexity. For
them, the paper gives a simple, yet detailed, exposition of an exponential lower bound on intuitionistic implicational
logic so that they cannot be fooled by the fact that Gordeev and Haeusler's claims have been published.
\item Researchers in proof complexity---not necessarily interested in Gordeev and Haeusler's claims---for whom the paper
brings a new, much shorter proof of the known implicational lower bound, bypassing implicational translation of full
intuitionistic logic. We stress that even though the proof system for which it is formulated is not traditional, it is
quite natural, and anyway the lower bound also applies to the standard Frege system for implicational
intuitionistic logic as the latter obviously embeds in dag-like natural deduction (up to subproofs of
Frege axioms, it can be thought of as natural deduction without the $\to$-introduction rule).
\end{itemize}

Our proof of the main lower bound does not involve any proof system other than dag-like natural deduction itself.
However, for the sake of completeness, we include an appendix showing the equivalence of dag-like natural deduction
with the standard intuitionistic implicational Frege system up to polynomial increase in proof size, as well as the
polynomial equivalence of both systems to their tree-like versions (adapting the original result of Kraj\'\i\v cek
along the lines of~\cite{ej:implic}). Thus, dag-like natural deduction does not offer any significant shortening of
proofs compared to the conventional tree-like natural deduction. The appendix may be of independent interest as we took
some effort to optimize the bounds.

An anonymous source pointed out that since Gordeev and Haeusler's ``horizontal compression'' only changes the shape of
the proof, but does not introduce any new formulas, their claims also contradict other well-known results in proof
complexity, namely constant-depth Frege lower bounds such as Beame et al.~\cite{bikppw:php}. For a sketch of the
argument, take a sequence of tautologies exponentially hard for constant-depth proofs, such as the pigeonhole
principle, and convert it to a sequence of (intuitionistically valid) implicational tautologies $\fii_n$ of polynomial
size and constant depth (measured, say, using the definition
$\operatorname{dp}(\fii\to\psi)=\max\{1+\operatorname{dp}(\fii),\operatorname{dp}(\psi)\}$). Each $\fii_n$ has a
cut-free sequent proof of polynomial height (and exponential size), which only involves formulas of constant depth by
the subformula property, and thus translates to a natural deduction proof of polynomial height using only formulas of
constant depth (with polynomially many distinct formulas). Gordeev and Haeusler's claims imply that this can be
compressed to a polynomial-size dag-like natural deduction proof using formulas of constant depth. The latter, however,
can be converted to a polynomial-size (classical) constant-depth sequent or Frege proof, contradicting the hardness of
the tautologies. We will not pursue this connection further in this paper, and leave the details to an interested
reader.

The paper is organized as follows. In Section~\ref{sec:preliminaries} we review the needed prerequisites such as
dag-like natural deduction and monotone Boolean circuits. Section~\ref{sec:lower-bound} is devoted to the proof of the
main exponential lower bound; we discuss extensions of the lower bound in Section~\ref{sec:extensions}, and we
conclude with a few remarks in Section~\ref{sec:conclusion}. We present the equivalence of dag-like natural deduction
to a Frege system in Appendix~\ref{sec:equiv-with-frege}, and the equivalence of both systems to their tree-like
versions in Appendix~\ref{sec:equivalence-dag-like}.

\section{Preliminaries}\label{sec:preliminaries}

The set $\Form$ of \emph{implicational formulas} (or just \emph{formulas} if no confusion arises) is the smallest set
that includes the set of \emph{propositional variables} (or \emph{atoms}) $\Var=\{p_n:n\in\omega\}$, and such that if
$\fii$ and~$\psi$ are formulas, then $(\fii\to\psi)$ is a formula. The \emph{size} $\lh\fii$ of a formula~$\fii$ is the
number of occurrences of variables and connectives in~$\fii$, i.e., $\lh{p_n}=1$ and
$\lh{(\fii\to\psi)}=1+\lh\fii+\lh\psi$. We may omit outer brackets in $\fii\to\psi$, and we treat $\to$ as a
right-associative operator so that, e.g., $\fii\to\psi\to\chi\to\omega$ stands for $(\fii\to(\psi\to(\chi\to\omega)))$.
(Despite these conventions, we may leave various redundant brackets in place to highlight the formula structure.)
We will denote formulas with lower-case Greek letters, and for convenience, we will often use lower-case Latin letters
(with indices and/or other decoration) other than $p_n$ for variables. We write $\vec p$ for a finite tuple of
variables $\p{p_i:i<n}$, especially if $n$ is immaterial; the notation $\fii(\vec p)$ indicates that all variables
occurring in~$\fii$ are among~$\vec p$.

Upper-case Greek letters will usually denote finite sets or sequences of formulas. Our indices generally start
from~$0$; in particular, $\p{\fii_i:i<n}$, or more concisely $\p{\fii_i}_{i<n}$, denotes the sequence
$\p{\fii_0,\dots,\fii_{n-1}}$ (which is the empty sequence $\p{}$ if $n=0$). The \emph{length} of a sequence
$\Gamma=\p{\fii_i}_{i<n}$, denoted $\lh\Gamma$, is~$n$, and the \emph{size} of~$\Gamma$, denoted $\dlh\Gamma$, is
$\sum_{i<n}\lh{\fii_i}$. If $\Gamma=\p{\fii_i}_{i<n}$ is a sequence
of formulas and $\psi\in\Form$, we introduce the abbreviation $\Gamma\to\psi$ for the formula%
\footnote{It might appear more visually pleasing to define it as $\fii_0\to\fii_1\to\dots\to\fii_{n-1}\to\psi$, but the
reverse order will be technically more convenient, e.g.\ in some inductive arguments in
Appendix~\ref{sec:equiv-with-frege}.}
\[\fii_{n-1}\to\dots\to\fii_1\to\fii_0\to\psi.\]
Formally, $\Gamma\to\psi$ is defined by induction on~$n$: $\p{\fii_i}_{i<0}\to\psi$ is $\psi$ and
$\p{\fii_i}_{i<n+1}\to\psi$ is $\fii_n\to\p{\fii_i}_{i<n}\to\psi$.

A \emph{substitution} is a mapping $\sigma\colon\Form\to\Form$ such that
$\sigma(\fii\to\psi)=(\sigma(\fii)\to\sigma(\psi))$ for all $\fii,\psi\in\Form$. If $\Gamma\sset\Form$, we write
$\sigma(\Gamma)=\{\sigma(\fii):\fii\in\Gamma\}$.

The \emph{intuitionistic implicational logic $\IPC_\to$} is defined by its consequence relation
${\vdash}\sset\pw\Form\times\Form$: we put $\Gamma\vdash\fii$ iff $\fii$ belongs to the smallest subset of $\Form$ that
is closed under the rule of modus ponens
\[\fii,\fii\to\psi\ru\psi,\]
and includes~$\Gamma$ and the logical axioms
\begin{align*}
\fii&\to\psi\to\fii\\
(\fii\to\psi\to\chi)&\to(\fii\to\psi)\to(\fii\to\chi)
\end{align*}
for $\fii,\psi,\chi\in\Form$. As is conventional, we omit braces around formulas on the left-hand side of~$\vdash$, and write
commas in place of $\cup$, so that, e.g., $\Gamma,\fii,\psi\vdash\chi$ stands for $\Gamma\cup\{\fii,\psi\}\vdash\chi$;
we may also coerce finite sequences $\Gamma$ to sets. We write $\vdash\fii$ for $\nul\vdash\fii$, in which case we say
that $\fii$ is an \emph{intuitionistic implicational tautology}, or \emph{$\IPC_\to$ tautology} for short.
\begin{Lem}[deduction theorem]\label{lem:ded}
Let $\Pi\sset\Form$, $\fii\in\Form$, and let $\Gamma$ be a finite sequence of formulas. Then
\[\Pi,\Gamma\vdash\fii\iff\Pi\vdash\Gamma\to\fii.\noproof\]
\end{Lem}

A \emph{Kripke model} is a structure $\p{W,{\le},{\model}}$, where $\le$ is a partial order on~$W$, and
${\model}\sset W\times\Form$ satisfies
\begin{align*}
x\model\fii&\,\implies\forall y\ge x\:y\model\fii,\\
x\model\fii\to\psi&\iff\forall y\ge x\:(y\model\fii\implies y\model\psi)
\end{align*}
for all $x\in W$ and $\fii,\psi\in\Form$. Unwinding the definitions, we see that for any sequence
$\Gamma=\p{\fii_i}_{i<n}$,
\[x\model\Gamma\to\psi\iff\forall y\ge x\:\bigl((\forall i<n\:y\model\fii_i)\implies y\model\psi\bigr).\]
A formula $\fii$ \emph{holds} in~$\p{W,{\le},{\model}}$ if $x\model\fii$ for all $x\in W$.

Intuitionistic logic is complete w.r.t.\ Kripke semantics, even if we only consider finite frames (see
e.g.~\cite{troe-vdal:constr,cha-zax}):
\begin{Thm}[finite model property]\label{thm:fmp}
A formula is an $\IPC_\to$ tautology if and only if it holds in all finite Kripke models.
\noproof\end{Thm}

Let us now present Gordeev and Haeusler's dag-like natural deduction
calculus{\def\thefootnote{\fnsymbol{footnote}}\footnotemark[2]} $\nmi$ based on~\cite{gor-hae:crap-ii}.
An \emph{$\nmi$-proof skeleton} is a finite directed acyclic graph (dag) $\p{V,E}$ with a unique node of
out-degree~$0$, called the \emph{root}, and with all nodes having in-degree at most~$2$; nodes of in-degree $0$, $1$,
and~$2$ are called leaves (assumptions), $\imi$-nodes, and $\ime$-nodes, respectively. If $\p{u,v}\in E$, then $u$ is a
\emph{premise}%
\footnote{In~\cite{gor-hae:crap-ii}, proofs go upside down so that edges are directed from conclusions to premises; we
reversed them to a more natural order. Also, they include an auxiliary repetition rule $\alpha/\alpha$
that we omit for simplicity (it can be eliminated from any $\nmi$ derivation without increasing its size).}%
{\def\thefootnote{\fnsymbol{footnote}}\footnotetext[2]{%
After this paper was published, Gordeev and Haeusler posted a new version of one of their preprints (\url{https://arxiv.org/abs/2311.17939v2})
where they claim that my results are ``wrong'' and ``misleading'' because I have ``misunderstood'' the definition of
$\nmi$ instead of a much more complicated definition of dag-like proof soundness involving $n$-ary repetition rules.
It is very difficult to characterize these claims as anything else than \emph{lying}. The reader is encouraged to
check for themselves that I have faithfully reproduced here the definition of $\nmi$ from~\cite[\S1]{gor-hae:crap-ii}
(Def.~1.1--L.~1.3) or \cite[\S2.1]{gor-hae:crap-pfth}, and that it is indeed this very system (rather than, say, the
subsequently introduced $\nmi^\flat$) that they claim to have polynomial-size proofs of all intuitionistic
implicational tautologies in~\cite[Cor.~2.7]{gor-hae:crap-ii} and \cite[Cor.~10]{gor-hae:crap-pfth}. I cannot be held
responsible for their claims being a moving target.}}
of~$v$. A \emph{thread} is a directed path starting from a leaf; a thread is \emph{maximal} if it ends in the root. An
\emph{$\nmi$-derivation $\p{V,E,\gamma}$} is an $\nmi$-proof skeleton $\p{V,E}$ endowed with a vertex labelling
$\gamma=\p{\gamma_v:v\in V}$ with $\gamma_v\in\Form$, such that for all $v\in V$:
\begin{itemize}
\item if $v$ is an $\imi$-node, it is labelled with an implication $\alpha\to\beta$ such that the premise of~$v$ is
labelled with~$\beta$;
\item if $v$ is an $\ime$-node, there are formulas $\alpha,\beta$ such that $v$ is labelled with~$\beta$, and the two
premises of~$v$ are labelled with $\alpha$ and $\alpha\to\beta$, respectively.
\end{itemize}
A thread with leaf~$v$ is \emph{discharged} if it contains an $\imi$-node labelled with $\alpha\to\beta$ where
$\alpha=\gamma_v$. Let $\fii\in\Form$ and $\Gamma\sset\Form$. An $\nmi$-derivation is an \emph{$\nmi$-derivation of
$\fii$ from~$\Gamma$} if the root is labelled~$\fii$ and the leaves of all undischarged maximal threads are labelled
with elements of~$\Gamma$. An \emph{$\nmi$-proof of~$\fii$} is an $\nmi$-derivation of $\fii$ from~$\nul$. The
\emph{number of lines} of an $\nmi$-derivation $\Pi=\p{V,E,\gamma}$ is $\lh V$, and the \emph{size} of~$\Pi$ is
$\dlh\Pi=\sum_{v\in V}\lh{\gamma_v}$.

It may be difficult to verify the condition on discharging maximal threads directly from the definition. As observed
in~\cite{gor-hae:crap-ii}, it can be checked efficiently as follows. Given an $\nmi$-derivation $\Pi=\p{V,E,\gamma}$,
we define for each $v\in V$ a set $A_v\sset\{\gamma_u:\text{$u$ is a leaf}\}$ by well-founded recursion:
\[A_v=\begin{cases}
\{\gamma_v\},&\text{$v$ is a leaf,}\\
A_u\bez\{\alpha\},&\text{$v$ is an $\imi$-node with premise~$u$ and $\gamma_v=\alpha\to\beta$,}\\
A_{u_0}\cup A_{u_1},&\text{$v$ is an $\ime$-node with premises $u_0$ and~$u_1$.}
\end{cases}\]
Note that given $\Pi$, we can compute $\p{A_v:v\in V}$ in polynomial time.
\begin{Lem}[\cite{gor-hae:crap-ii}]\label{lem:A}
An $\nmi$-derivation $\p{V,E,\gamma}$ with root~$\roo$ is a derivation of $\gamma_\roo$ from $\Gamma$ if and only if
$A_\roo\sset\Gamma$.
\end{Lem}
\begin{Pf}
Show that $A_v$ is the set of labels of undischarged threads ending in~$v$ by well-founded induction on~$v$.
\end{Pf}

Likewise, we can show the soundness of $\nmi$-derivations by well-founded induction on~$v$, using the deduction theorem:
\begin{Lem}\label{cor:A}
For any $\nmi$-derivation $\p{V,E,\gamma}$ and $v\in V$, $A_v\vdash\gamma_v$.
\noproof\end{Lem}

On the other hand, \emph{tree-like} $\nmi$~derivations are the same as the implicational fragment of the usual
Gentzen--Prawitz natural deduction (see e.g.\ \cite{praw:nd,man-gal-zach}). This implies the completeness of the
calculus, as observed in~\cite{gor-hae:crap-ii}:
\begin{Lem}\label{lem:nmi-comp}
A formula~$\fii$ is an $\IPC_\to$ tautology if and only if it has an $\nmi$-proof.
\noproof\end{Lem}

We assume familiarity with classical propositional logic, but briefly, we consider formulas built from propositional
variables using the connectives $\{\to,\land,\lor,\neg,\top,\bot\}$. An \emph{assignment} to a set of variables~$X$ is
a function $a\colon X\to\two$, where $\two=\{0,1\}$. We denote the set of all such assignments as $\two^X$. For any
$a\in\two^X$ and a formula $\fii$ over variables~$X$, we define the relation $a\model\fii$ (in words, $a$
\emph{satisfies} $\fii$) in the usual way:
\begin{align*}
a\model p&\iff a(p)=1,\quad p\in X,\\
a\model(\fii\to\psi)&\iff a\nmodel\fii\text{ or }a\model\psi,\\
a\model\neg\fii&\iff a\nmodel\fii,
\end{align*}
and so on for the other connectives. A formula $\fii$ is a \emph{classical tautology} if $a\model\fii$ for all
assignments $a$ to the variables of~$\fii$.

We also need a bit of circuit complexity. A \emph{monotone circuit} over a set~$X$ of \emph{variables} is $C=\p{V,E,g}$
where $\p{V,E}$ is a dag with a unique node $\roo$ of out-degree~$0$ (the \emph{root}), endowed with a labelling
$g\colon V\to X\cup\{\land,\lor\}$ such that nodes $v$ with $g(v)\in X$ have in-degree~$0$. Nodes $v\in V$ are also
called \emph{gates}, and edges $e\in E$ are called \emph{wires}. We may write $C(\vec p)$ to denote that $C$ is a
circuit over a finite tuple of variables~$\vec p$. The \emph{size} of a circuit $C=\p{V,E,g}$ is $\lh C=\lh E$ (i.e.,
the number of wires). By well-founded recursion, any assignment $a\colon X\to\two$ extends to a unique
function $\hat a\colon V\to\two$, called the \emph{evaluation} of~$C$, such that
\[\hat a(v)=\begin{cases}
a(g(v)),&g(v)\in X,\\
\rlap{$\inf$}\hphantom\sup\{\hat a(u):\p{u,v}\in E\},&g(v)=\land,\\
\sup\{\hat a(u):\p{u,v}\in E\},&g(v)=\lor,
\end{cases}\]
where $\inf\nul=1$, $\sup\nul=0$ (thus $\land$- and $\lor$-gates without inputs act as constants $\top$ and~$\bot$,
respectively). A circuit~$C$ with root~$\roo$ \emph{computes} a Boolean function $f\colon\two^X\to\two$ if
$f(a)=\hat a(\roo)$ for each $a\in\two^X$. More generally, a \emph{disjoint pair} is $P=\p{P^0,P^1}$ where
$P^0,P^1\sset\two^X$ and $P^0\cap P^1=\nul$; a circuit~$C$ \emph{separates~$P$} if $\hat a(\roo)=i$ for each $i\in\two$
and $a\in P^i$. We will write $a\model C$ for $\hat a(\roo)=1$.

Let $\vec p$, $\vec q$, and $\vec r$ be pairwise disjoint tuples of variables, and $\fii(\vec p,\vec q)$
and~$\psi(\vec p,\vec r)$ classical formulas. Then a circuit~$C(\vec p)$ \emph{interpolates} the implication
$\fii\to\psi$ (which must be a classical tautology) if $\fii(\vec p,\vec q)\to C(\vec p)$ and
$C(\vec p)\to\psi(\vec p,\vec r)$ are classical tautologies (i.e., $a\model\fii\implies a\model C$ and $a\model
C\implies a\model\psi$ for all assignments $a\in\two^{\{\vec p,\vec
q,\vec r\}}$), or
in other words, if $C$ separates the \emph{interpolation pair} $\itp_{\fii,\psi}=\p{\itp_\psi^0,\itp_\fii^1}$, where
\begin{align*}
\itp_\psi^0&=\{a\in\two^{\vec p}:\exists c\in\two^{\vec r}\:\p{a,c}\nmodel\psi\},\\
\itp_\fii^1&=\{a\in\two^{\vec p}:\exists b\in\two^{\vec q}\:\p{a,b}\model\fii\}.
\end{align*}

For any $n\ge2$, the \emph{Colouring--Cocolouring} disjoint pair $\ccp_n=\p{\ccp_n^0,\ccp_n^1}$
over the set of variables $X_n=\binom{[n]}2$ (i.e., the set of unordered pairs of elements of $[n]=\{0,\dots,n-1\}$) is
defined by
\begin{align*}
\ccp_n^0&=\bigl\{E\sset X_n:\text{the graph $\p{[n],E}$ is $k$-colourable}\bigr\},\\
\ccp_n^1&=\bigl\{E\sset X_n:\text{the graph $\p{[n],\ob E}$ is $k$-colourable}\bigr\},
\end{align*}
where $\ob E=X_n\bez E$, $k=\cl{\sqrt n}-1$, and we identify $E\sset X_n$ with its characteristic function
$X_n\to\two$. To see that $\ccp_n^0\cap\ccp_n^1=\nul$, observe that if $c_0,c_1\colon[n]\to[k]$ are $k$-colourings of
$\p{[n],E}$ and $\p{[n],\ob E}$, respectively, then $c_0\times c_1\colon[n]\to[k]\times[k]$ is an injection, thus $n\le
k^2$.

An exponential lower bound on the monotone circuit complexity (and even monotone \emph{real} circuit complexity)
of~$\ccp_n$ was proved by Hrube\v s and Pudl\'ak~\cite[Thm.~10]{hru-pud:col-col}, using machinery from
Jukna~\cite{juk:boolf}:
\begin{Thm}\label{thm:hru-pud}
For $n\gg0$, all monotone circuits separating $\ccp_n$ have size $2^{\Omega(k^{1/4})}=2^{\Omega(n^{1/8})}$.
\noproof\end{Thm}

Strictly speaking, Hrube\v s and Pudl\'ak work with \emph{bounded fan-in} monotone circuits, i.e., such that the
in-degree of all gates is at most~$2$, and they measure size by the number of gates. This makes no difference, as a
$d$-ary $\land$- or $\lor$-gate can be simulated by $d-1$ binary gates using $2(d-1)$~wires, thus any monotone circuit
with $s$~wires can be transformed to a bounded fan-in monotone circuit with $s'\le2s$ wires; moreover, a circuit with
$s'$~wires has at most $s'+1$ gates (we may associate each node other than the root with an outgoing wire). This mild
size increase does not affect the shape of the lower bound in Theorem~\ref{thm:hru-pud}.

\section{An exponential lower bound}\label{sec:lower-bound}

In this section, we will prove our main lower bound, viz.\ there is an explicit sequence of implicational
intuitionistic tautologies that require $\nmi$-proofs with exponentially many lines.

Let us start with construction of the $\IPC_\to$ tautologies, which will express the disjointness of~$\ccp_n$.
Intuitionistic tautologies expressing disjointness of the Clique--Colouring pair were first considered by Hrube\v
s~\cite{hru:lbint}; they were made negation-free in Je\v r\'abek~\cite{ej:sfef}, and implicational in Je\v
r\'abek~\cite{ej:implic}. We will further simplify the tautologies from~\cite{ej:implic} by using a somewhat more
direct translation to implicational logic, and by employing the Colouring--Coclouring pair in place of
Clique--Colouring, which leads to more symmetric (and shorter) formulas. Fix $n\ge2$ and $k=\cl{\sqrt n}-1$. Our
tautologies will employ variables $p_{ij}$ and~$p'_{i,j}$ ($i<j<n$), representing the edge relation of a graph
$G=\p{[n],E}$ and its complement, and variables $q_{il}$ and $r_{il}$ ($i<n$, $l<k$),
representing a $k$-colouring of~$G$ and of its complement (respectively).

To motivate the formal definition below, we can state in classical propositional logic that $\vec q$ define a (possibly
multivalued) $k$-colouring of~$G$ by the formula
\[\ET_{i<n}\LOR_{l<k}q_{il}\land\ET_{\substack{\mclap{i<j<n}\\l<k}}\neg(q_{il}\land q_{jl}\land p_{ij}),\]
and similarly for the complement, thus the disjointness of $\ccp_n$ is expressed by the classical tautology
\[\Bigl(\ET_{i<n}\LOR_{l<k}q_{il}\to\LOR_{\substack{\mclap{i<j<n}\\l<k}}(q_{il}\land q_{jl}\land p_{ij})\Bigr)
  \lor\Bigl(\ET_{i<n}\LOR_{l<k}r_{il}\to\LOR_{\substack{\mclap{i<j<n}\\l<k}}(r_{il}\land r_{jl}\land\neg p_{ij})\Bigr),\]
which can be made negation-free using the $\vec p\,'$ variables:
\[\ET_{i<j<n}(p_{ij}\lor p'_{ij})\to
  \Bigl(\ET_{i<n}\LOR_{l<k}q_{il}\to\LOR_{\substack{\mclap{i<j<n}\\l<k}}(q_{il}\land q_{jl}\land p_{ij})\Bigr)
  \lor\Bigl(\ET_{i<n}\LOR_{l<k}r_{il}\to\LOR_{\substack{\mclap{i<j<n}\\l<k}}(r_{il}\land r_{jl}\land p'_{ij})\Bigr).\]
This turns out to be an intuitionistic tautology as well. In order to convert it to an implicational tautology, we
introduce further auxiliary variables $u$, $v$, and~$w$: the idea is to rewrite an implication $\psi\to\chi$ as
$(\chi\to u)\to(\psi\to u)$, where $\psi\to u$ and $\chi\to u$ can be written using implicational formulas when $\psi$
and $\chi$ are monotone formulas. After some manipulation we end up with the following:
\begin{Def}\label{def:taut}
Let $n\ge2$ and $k=\cl{\sqrt n}-1$. We define the following implicational formulas in variables $p_{ij}$, $p'_{ij}$,
$q_{il}$, $r_{il}$, $u$, $v$, and~$w$, where $i<j<n$ and~$l<k$:
\begin{align*}
\alpha_n(\vec p,\vec q,v)
&=\pp[big]{\p{q_{il}\to v}_{l<k}\to v}_{i<n}\to\p{q_{il}\to q_{jl}\to p_{ij}\to v}_{\substack{i<j<n\\l<k}}\to v,\\
\tau_n(\vec p,\vec p\,',\vec q,\vec r,u,v,w)
&=\pp[big]{(p_{ij}\to u)\to(p'_{ij}\to u)\to u}_{i<j<n}\\
&\qquad\qquad\to\bigl(\alpha_n(\vec p,\vec q,v)\to u\bigr)\to\bigl(\alpha_n(\vec p\,',\vec r,w)\to u\bigr)\to u.
\end{align*}
(The order in which we enumerate the multiply-indexed sequences such as $\p\dots_{i<j<n}$ does not matter.)
\end{Def}
\begin{Obs}\label{obs:size}
$\lh{\tau_n}=O(n^2k)=O(n^{5/2})$.
\noproof\end{Obs}

\begin{Lem}\label{lem:taut}
The formulas $\tau_n$ are intuitionistic implicational tautologies.
\end{Lem}
\begin{Pf}
Assume for contradiction that $\tau_n$ does not hold in a finite Kripke model $\p{W,{\le},{\model}}$. This means that
there exists $x\in W$ such that $x\model(p_{ij}\to u)\to(p'_{ij}\to u)\to u$ for all $i<j<n$,
$x\model\alpha_n(\vec p,\vec q,v)\to u$, $x\model\alpha_n(\vec p\,',\vec r,w)\to u$, but $x\nmodel u$. Replacing $x$
with some $\tilde x\ge x$ if necessary, we may assume that $x$ is maximal such that $x\nmodel u$, i.e., $x'\model u$
for all $x'>x$.

For each $i<j<n$, $x\model(p_{ij}\to u)\to(p'_{ij}\to u)\to u$ implies that $x\nmodel p_{ij}\to u$ or
$x\nmodel p'_{ij}\to u$. Since $u$ is true in all $x'>x$, we obtain
\begin{equation}\label{eq:1}
\forall i<j<n\:(x\model p_{ij}\text{ or }x\model p'_{ij}).
\end{equation}

Since $x\model\alpha_n(\vec p,\vec q,v)\to u$, we have $x\nmodel\alpha_n(\vec p,\vec q,v)$, thus there exists $y\ge x$
such that $y\model\p{q_{il}\to v}_{l<k}\to v$ for all $i<n$, and $y\model q_{il}\to q_{jl}\to p_{ij}\to v$ for all
$i<j<n$ and $l<k$, but $y\nmodel v$. As above, we may assume that $y'\model v$ for all $y'>y$. Then for every $i<n$,
$y\model\p{q_{il}\to v}_{l<k}\to v$ implies $y\nmodel q_{il}\to v$ for some $l<k$, whence $y\model q_{il}$ by
maximality. That is, we can find a colouring function $c\colon[n]\to[k]$ such that $y\model q_{i,c(i)}$ for all $i<n$.

If $i<j<n$ are such that $c(i)=c(j)=l$, then $y\model q_{il}\to q_{jl}\to p_{ij}\to v$ and $y\nmodel v$ implies
$y\nmodel p_{ij}$, and a fortiori $x\nmodel p_{ij}$. This shows that $c$ is a proper $k$-colouring of the graph
$\p{[n],E}$, where $E=\bigl\{\{i,j\}:x\model p_{ij}\bigr\}$.

Since $x\model\alpha_n(\vec p\,',\vec r,w)\to u$, the same argument gives a $k$-colouring $c'\colon[n]\to[k]$ of
$\p{[n],E'}$, where $E'=\bigl\{\{i,j\}:x\model p'_{ij}\bigr\}$. But then \eqref{eq:1} implies that the function
$c\times c'\colon[n]\to[k]\times[k]$ is injective, thus $n\le k^2<n$, a contradiction.
\end{Pf}

The remaining task is to prove a form of monotone feasible interpolation (based on feasible disjunction property) for
$\nmi$, which will imply an exponential lower bound for the $\tau_n$~tautologies using Theorem~\ref{thm:hru-pud}. There are
many ways how to prove the disjunction property of intuitionistic logic and various intuitionistic theories, one of
them being \emph{Kleene's slash} \cite{kle:sls}. Efficient versions of Kleene's slash were used by Ferrari, Fiorentini,
and Fiorino~\cite{fff} (under the umbrella machinery of ``extraction calculi'') to prove the feasible disjunction
property for the intuitionistic natural deduction system (which was originally proved by Buss and
Mints~\cite{buss-mints} using a form of cut elimination); by Mints and Kojevnikov~\cite{min-koj} to prove the
polynomial equivalence of intuitionistic Frege systems using admissible rules (with a considerably simplified argument
given by Je\v r\'abek~\cite{ej:modfrege}); and by Je\v r\'abek~\cite{ej:sfef} to prove an exponential lower bound on
intuitionistic Extended Frege proofs. We will adapt the argument from~\cite{ej:sfef} to a purely implicational setting,
using a disjunction-free analogue of the disjunction property.
\begin{Def}\label{def:sls}
If $P\sset\Form$, a \emph{$P$-slash} is a unary predicate $\sls$ on~$\Form$ such that
\begin{align*}
\sls(\fii\to\psi)&\iff\bigl(\Sls\fii\implies\sls\psi\bigr)\\
\intertext{for all $\fii,\psi\in\Form$, where we define the short-hand}
\Sls\fii&\iff\sls\fii\text{ and }\fii\in P.
\end{align*}
If $\Gamma$ is a set of formulas, we write $\Sls\Gamma$ if $\Sls\fii$ for all $\fii\in\Gamma$. When we need to consider
several slash operators at the same time, we may distinguish them by subscripts, which are carried over to~$\Sls$. We
warn the reader that a $P$-slash is not uniquely determined by~$P$, as we have liberty in defining $\sls p$ for
$p\in\Var$; however, an arbitrary choice of $|$ on~$\Var$ has a unique extension to a $P$-slash.

If $\Pi=\p{V,E,\gamma}$ is an $\nmi$-derivation, a set $P\sset\Form$ is \emph{$\Pi$-closed} if $A_v\sset
P\implies\gamma_v\in P$ for all $v\in V$.
\end{Def}

Unwinding the definition, we obtain:
\begin{Obs}\label{obs:sls-seq}
If $\Gamma$ is a finite sequence of formulas, and $\fii\in\Form$, then
\[\sls(\Gamma\to\fii)\iff\bigl(\Sls\Gamma\implies\sls\fii\bigr).\noproof\]
\end{Obs}

We first verify that being $\Pi$-closed is enough to ensure the soundness of the slash:
\begin{Lem}\label{lem:sls-sound}
Let $\Pi$ be an $\nmi$-proof of~$\fii$, $P$ be a $\Pi$-closed set of formulas, and $\sls$ be a $P$-slash. Then $\Sls\fii$.
\end{Lem}
\begin{Pf}
We prove
\begin{equation}\label{eq:3}
\Sls A_v\implies\Sls\gamma_v
\end{equation}
by well-founded induction on~$v\in V$. This is trivial if $v$ is a leaf. Let $v$ be an $\ime$-node with premises
$u_0$, $u_1$, such that $\gamma_{u_0}=\alpha$, $\gamma_{u_1}=(\alpha\to\beta)$, and $\gamma_v=\beta$, and assume
$\Sls A_v$. Since $A_{u_i}\sset A_v$, the induction hypothesis gives $\Sls\alpha$ and $\Sls(\alpha\to\beta)$. Then the
definition of $\sls(\alpha\to\beta)$ ensures $\sls\beta$, and $A_v\sset P$ implies $\beta\in P$ as $P$ is
$\Pi$-closed, thus $\Sls\beta$.

Finally, let $v$ be an $\imi$-node with premise~$u$ such that $\gamma_u=\beta$ and $\gamma_v=(\alpha\to\beta)$, and
assume $\Sls A_v$. Then $A_v\sset P$ implies $\gamma_v\in P$ as $P$ is $\Pi$-closed, hence it suffices to show
$\sls(\alpha\to\beta)$. Thus, assume $\Sls\alpha$; since $A_u\sset A_v\cup\{\alpha\}$, we have $\Sls A_u$, thus $\Sls\beta$ by
the induction hypothesis.
\end{Pf}

Next, we need to furnish ourselves with $\Pi$-closed sets.
\begin{Def}\label{def:cl}
Let $\Pi=\p{V,E,\gamma}$ be an $\nmi$-derivation and $P\sset\Form$. The \emph{$\Pi$-closure of~$P$}, denoted
$\cls_\Pi(P)$, is $P_{\lh V}$, where we define $P_i$ for each $i\in\omega$ by
\begin{align*}
P_0&=P,\\
P_{i+1}&=P_i\cup\{\gamma_v:v\in V,A_v\sset P_i\}.
\end{align*}
\end{Def}
\begin{Lem}\label{lem:cl}
Let $\Pi$ be an $\nmi$-derivation and $P\sset\Form$.
\begin{enumerate}
\item\label{item:1} The set $\cls_\Pi(P)\supseteq P$ is $\Pi$-closed.
\item\label{item:2} $P\vdash\fii$ for all $\fii\in\cls_\Pi(P)$.
\end{enumerate}
\end{Lem}
\begin{Pf}

\ref{item:1}: Let $\Pi=\p{V,E,\gamma}$ and $t=\lh V$. It is clear from the definition that if $P_i=P_{i+1}$, then $P_i$
is $\Pi$-closed, and $P_i=P_j$ for all $j\ge i$. Thus, it suffices to shows that $P_i=P_{i+1}$ for some $i\le t$. If
not, then $P=P_0\ssset P_1\ssset\dots\ssset P_{t+1}$, thus $\lh{P_i\bez P}\ge i$ for each $i\le t+1$ by induction
on~$i$; but $P_i\sset P\cup\{\gamma_v:v\in V\}$, thus $t\ge\lh{P_{t+1}\bez P}\ge t+1$, a contradiction.

\ref{item:2}: We can prove $P\vdash\fii$ for all $\fii\in P_i$ by induction on~$i$ using Corollary~\ref{cor:A}.
\end{Pf}

It will be crucial in what follows that $\Pi$-closure is efficiently computable: e.g., it is easy to see that it is
computable in polynomial time; but what we will actually need is that it is computable by polynomial-size monotone
circuits in the following sense:
\begin{Lem}\label{lem:cl-circ}
Let $\Pi=\p{V,E,\gamma}$ be an $\nmi$-derivation with $t=\lh V$ lines, $F=\{\fii_i:i<n\}\sset\Form$ be such that
$\{\gamma_v:v\in V\}\sset F$, and $\fii\in F$.

Then there exists a monotone circuit $C$ of size $O(t^3)$ over variables $X=\{x_i:i<n\}$ such that for every
assignment $a\in\two^X$,
\[a\model C\iff\fii\in\cls_\Pi\bigl(\{\fii_i:a(x_i)=1\}\bigr).\]
\end{Lem}
\begin{Pf}
We may assume $\fii=\fii_0$. If $\fii\notin F_\Pi=\{\gamma_v:v\in V\}$, then $\fii\in\cls_\Pi(P)\iff\fii\in P$, which
is computable by the trivial circuit $C=x_0$, thus we may assume $\fii\in F_\Pi$. More generally, we observe that
$\cls_\Pi(P)=P\cup\cls_\Pi(P\cap F_\Pi)$, thus we may assume $F=F_\Pi$; in particular, $n\le t$.

We consider a circuit~$C$ with nodes $y_{i,j}$ for $i<n$ and $j\le t$, and $z_{v,j}$ for $v\in V$ and $j<t$, wired such
that
\begin{align*}
y_{i,0}&\equiv x_i,\\
y_{i,j+1}&\equiv y_{i,j}\lor\LOR_{\substack{v\in V\\\gamma_v=\fii_i}}z_{v,j},\\
z_{v,j}&\equiv\ET_{\substack{i<n\\\fii_i\in A_v}}y_{i,j}.
\end{align*}
We define the root of~$C$ to be $y_{0,t}$ (and we remove nodes from which $y_{0,t}$ is not reachable to satisfy the
formal definition of a circuit). It follows from the definition by induction on~$j$ that if $a\in\two^X$ and
$P=\{\fii_i:a(x_i)=1\}$, then
\begin{align*}
\hat a(y_{i,j})=1&\iff\fii_i\in P_j,\\
\hat a(z_{v,j})=1&\iff A_v\sset P_j,
\end{align*}
where $\hat a$ is the evaluation of~$C$ extending~$a$. Consequently, $\fii\in\cls_\Pi(P)\iff\ob a(y_{0,t})=1$.

In order to determine $\lh C$, for each $j<t$ there are $n$~wires going from $y_{i,j}$ to~$y_{i,j+1}$, $t$~wires (one
for each $v\in V$) going from $z_{v,j}$ to $y_{i,j+1}$ where $\gamma_v=\fii_i$, and $\sum_v\lh{A_v}\le nt$ wires going
from $y_{i,j}$ to $z_{v,j}$ such that $\fii_i\in A_v$. Thus, $\lh C\le(n+t+nt)t=O(nt^2)=O(t^3)$, using $n\le t$.
\end{Pf}

Pudl\'ak~\cite{pud:fi} showed that the feasible disjunction property of intuitionistic calculi can serve a similar role
as feasible interpolation for classical proof systems, and as such implies conditional lower bounds on the length of
intuitionistic proofs. Hrube\v s~\cite{hru:lbint} discovered how to modify the set-up to obtain an analogue of feasible
\emph{monotone} interpolation (first considered by Kraj\'\i\v cek~\cite{kra:fi}), which yields unconditional
exponential lower bounds utilizing monotone circuit lower bounds such as Alon and Boppana~\cite{alonbop}. These results
naturally rely on the presence of disjunction. Je\v r\'abek~\cite{ej:implic} obtained a lower bound on implicational
intuitionistic logic based using implicational translations of intuitionistic formulas, but here we follow a more
direct approach: we introduce a version of feasible monotone interpolation based on a ``disjunction-free disjunction
property''. This is the main new idea of this paper. To help the reader with intuition, we first prove a most simple
version of disjunction-free feasible disjunction property%
\footnote{It is not surprising that $\alpha_0\lor\alpha_1$ can be expressed by an implicational formula as in
Lemma~\ref{lem:fdp-imp}; what is supposed to be novel here is the way to prove the feasible disjunction property for this
formulation without reintroducing disjunctions.}%
, although we will not really use this statement later.
\begin{Lem}\label{lem:fdp-imp}
Given an $\nmi$-proof $\Pi$ of a formula~$\fii$ of the form
\[(\alpha_0\to u)\to(\alpha_1\to u)\to u,\]
where the variable~$u$ does not occur in $\alpha_0$ and~$\alpha_1$, we can compute in polynomial time an $i\in\{0,1\}$
such that $\vdash\alpha_i$.
\end{Lem}
\begin{Pf}
Put $P=\cls_\Pi(\alpha_0\to u,\alpha_1\to u)$, and let $\sls$ be a $P$-slash such that $\nsls u$. Since $\sls\fii$ by
Lemmas \ref{lem:sls-sound} and~\ref{lem:cl}, we have $\nSls(\alpha_i\to u)$ for some~$i<2$ by Observation~\ref{obs:sls-seq}. In view of
$(\alpha_i\to u)\in P$, this means $\nsls(\alpha_i\to u)$, thus $\Sls\alpha_i$. That is, we have verified
\[\alpha_0\in P\text{ or }\alpha_1\in P.\]
Given~$\Pi$, we can compute $P$ in polynomial time, hence we can compute $i<2$ such that $\alpha_i\in P$. It remains to
verify that this implies $\vdash\alpha_i$. Lemma~\ref{lem:cl} gives
\[\alpha_0\to u,\alpha_1\to u\vdash\alpha_i.\]
But $u$ does not occur in $\alpha_i$, hence we may substitute it with~$\top$, obtaining $\vdash\alpha_i$.
\end{Pf}

We now generalize this argument to a Hrube\v s-style feasible monotone interpolation.
\begin{Thm}\label{thm:fmi-imp}
Let $\vec p=\p{p_i:i<n}$, $\vec p\,'=\p{p'_i:i<n}$, $\vec q$, $\vec r$, and $u$ be pairwise disjoint tuples of
variables, and assume that a formula~$\fii$ of the form
\[\pp[big]{(p_i\to u)\to(p'_i\to u)\to u}_{i<n}
  \to\bigl(\alpha_0(\vec p,\vec q)\to u\bigr)\to\bigl(\alpha_1(\vec p\,',\vec r)\to u\bigr)\to u\]
has an $\nmi$-proof with $t$~lines. Then there exists a monotone circuit $C(\vec p)$ of size $O(t^3)$
that interpolates the classical tautology
\[\neg\alpha_1(\neg\vec p,\vec r)\to\alpha_0(\vec p,\vec q),\]
where $\neg\vec p$ denotes $\p{\neg p_i:i<n}$.
\end{Thm}
\begin{Pf}
Let $\Pi=\p{V,E,\gamma}$ be a proof of~$\fii$ with $s$ lines. If $I\sset[n]$, we write $p^{}_I=\{p_i:i\in I\}$, and
similarly for~$p'_I$. We define
\begin{align*}
P&=\{(p_i\to u)\to(p'_i\to u)\to u:i<n\}\cup\{\alpha_j\to u:j<2\},\\
P_{I,J}&=\cls_\Pi(P\cup p^{}_I\cup p'_J)
\end{align*}
for each $I,J\sset[n]$. Let $\sls_{I,J}$ be a $P_{I,J}$-slash such that $\nsls_{I,J}u$ and $\sls_{I,J}x$ for all
variables $x\ne u$.

If $i\in I$, then $\Sls_{I,J}p_i$, thus $\nsls_{I,J}(p_i\to u)$, and $\sls_{I,J}(p_i\to u)\to(p'_i\to u)\to
u$ by Observation~\ref{obs:sls-seq}. Likewise if $i\in J$, using $\nsls_{I,J}(p'_i\to u)$. In view of $(p_i\to
u)\to(p'_i\to u)\to u \in P_{I,J}$, we obtain
\[I\cup J=[n]\implies\Sls_{I,J}\bigl\{(p_i\to u)\to(p'_i\to u)\to u:i<n\bigr\}.\] On the other hand, $\sls_{I,J}\fii$
by Lemmas \ref{lem:sls-sound} and~\ref{lem:cl}, thus assuming $I\cup J=[n]$, Observation~\ref{obs:sls-seq} implies
$\nSls_{I,J}(\alpha_j\to u)$ for some $j<2$. Since $\alpha_j\to u$ is in~$P_{I,J}$, this means
$\nsls_{I,J}(\alpha_j\to u)$, which implies $\Sls_{I,J}\alpha_j$. That is,
\[I\cup J=[n]\implies\alpha_0\in P_{I,J}\text{ or }\alpha_1\in P_{I,J}.\]
Applying this to $J=\ob I:=[n]\bez I$, and using the monotonicity of $\cls_\Pi$, we obtain
\begin{equation}\label{eq:4}
\forall I\sset[n]\:\bigl(\alpha_0\in P_{I,[n]}\text{ or }\alpha_1\in P_{[n],\ob I}\bigr).
\end{equation}

Put $F=P\cup p^{}_{[n]}\cup p'_{[n]}\cup\{\gamma_v:v\in V\}$. By Lemma~\ref{lem:cl-circ}, there is a monotone circuit of
size $O(t^3)$ that determines whether $\alpha\in\cls_\Pi(S)$ for a given $S\sset F$, using variables corresponding to
each $p_i\in F$, which we may identify with $p_i$ itself, variables corresponding to formulas in $P\cup p'_{[n]}$,
which we may substitute with~$\top$, and variables corresponding to other formulas from~$F$, which we may substitute
with~$\bot$. We obtain a monotone circuit $C(\vec p)$ of size $O(t^3)$ such that
\begin{equation}\label{eq:5}
a\model C\iff\alpha\in P_{I(a),[n]}
\end{equation}
for all assignments~$a$, where $I(a)=\{i<n:a(p_i)=1\}$.

We claim that $C$ interpolates $\neg\alpha_1(\neg\vec p,\vec r)\to\alpha_0(\vec p,\vec q)$. Let $a\in\two^{\{\vec p,\vec
q,\vec r\}}$. On the one hand, assume $a\model C$; we need to show $a\model\alpha_0$. We have $\alpha_0\in P_{I(a),[n]}$
by~\eqref{eq:5}. Since all formulas in~$P$ are implied by~$u$, we have
\[p^{}_{I(a)},p'_{[n]},u\vdash\alpha_0(\vec p,\vec q)\]
by Lemma~\ref{lem:cl}. But $\alpha_0$ does not contain the variables $p'_i$ or~$u$, hence we may substitute these
with~$\top$, obtaining
\[p^{}_{I(a)}\vdash\alpha_0(\vec p,\vec q).\]
Since $a\model p^{}_{I(a)}$, also $a\model\alpha_0$.

On the other hand, assume $a\nmodel C$; we will verify $a\model\alpha_1(\neg\vec p,\vec r)$. We have $\alpha_0\notin
P_{I(a),[n]}$ by~\eqref{eq:5}, hence $\alpha_1\in P_{[n],\ob{I(a)}}$ by~\eqref{eq:4}, thus
\[p^{}_{[n]},p'_{\ob{I(a)}},u\vdash\alpha_1(\vec p\,',\vec r)\]
by Lemma~\ref{lem:cl}. Substituting $\top$ for $\vec p$ and~$u$, we obtain
\[p'_{\ob{I(a)}}\vdash\alpha_1(\vec p\,',\vec r).\]
Finally, we can substitute $p'_i$ with $\neg p_i$ for each~$i$, getting
\[\neg p^{}_{\ob{I(a)}}\vdash\alpha_1(\neg\vec p,\vec r)\]
(in intuitionistic or classical logic with~$\neg$). Since $a$ satisfies the left-hand side, this implies
$a\model\alpha_1(\neg\vec p,\vec r)$.
\end{Pf}

We are ready to prove the main lower bound by applying Theorem~\ref{thm:fmi-imp} to the $\tau_n$~tautologies from
Definition~\ref{def:taut}; we only need to observe that interpolation of the implication $\neg\alpha_n(\neg\vec p,\vec
r,w)\to\alpha_n(\vec p,\vec q,v)$ is essentially identical to separation of the $\ccp_n$ disjoint pair.
\begin{Thm}\label{thm:main}
If $n$ is sufficiently large, then every $\nmi$-proof of $\tau_n$ has at least $2^{\Omega(n^{1/8})}$ lines.

Consequently, there are infinitely many intuitionistic implicational tautologies~$\fii$ such that every $\nmi$-proof
of~$\fii$ needs to have at least $2^{\Omega(\lh\fii^{1/20})}$ lines.
\end{Thm}
\begin{Pf}
It suffices to prove the first part; the second part then follows using Observation~\ref{obs:size}.

If $\tau_n$ has an $\nmi$-proof with $t$~lines, there is a monotone circuit $C(\vec p)$ of size $O(t^3)$ that
interpolates
\begin{equation}\label{eq:6}
\neg\alpha_n(\neg\vec p,\vec r,w)\to\alpha_n(\vec p,\vec q,v)
\end{equation}
by Theorem~\ref{thm:fmi-imp}. We claim that $C$ separates $\ccp_n$, which implies $t=2^{\Omega(n^{1/8})}$ by
Theorem~\ref{thm:hru-pud}.

Let $E\sset\binom{[n]}2$, and let $e$ be the corresponding assignment to~$\vec p$, i.e., for each $i<j<n$,
\[e(p_{ij})=1\iff\{i,j\}\in E.\]
Assume that $\p{[n],E}$ is $k$-colourable; we need to show $e\nmodel C$. Fix a vertex colouring
$c\colon[n]\to[k]$, and extend $e$ to an assignment on $\vec q$ and~$v$ by $e(v)=0$ and
\[e(q_{il})=1\iff c(i)=l\]
for each $i<n$ and~$l<k$. Then for every $i<n$, $e\nmodel q_{i,c(i)}\to v$, thus $e\model\p{q_{il}\to v}_{l<k}\to v$.
Likewise, for every $i<j<n$ and $l<k$, $e\model q_{il}\to q_{jl}\to p_{ij}\to v$, i.e.,
$e\nmodel q_{il}\land q_{jl}\land p_{ij}$: if $c(i)=l=c(j)$, then $\{i,j\}\notin E$ as $c$ is a proper colouring. Thus,
$e\nmodel\alpha_n(\vec p,\vec q,v)$, which implies $e\nmodel C(\vec p)$ as $C$ interpolates~\eqref{eq:6}.

A symmetrical argument shows that if $\p{[n],\ob E}$ is $k$-colourable, then $e$ extends to an assignment such that
$e\nmodel\alpha_n(\neg\vec p,\vec r,w)$, whence $e\model C(\vec p)$.
\end{Pf}

\section{Extensions}\label{sec:extensions}

The goal of the previous section has been to get to the basic lower bound (Theorem~\ref{thm:main}) as directly and as simply
as possible. However, if we expend more effort, we can improve the result in various ways---more or less up to the
strength of Theorem~4.22 of~\cite{ej:implic}. We briefly indicate these modifications and their difficulty below, but
we omit most details, and keep this section informal, as it is essentially an extended remark. We refer the reader to
\cite{ej:sfef,ej:implic} for missing definitions.

\paragraph{Logics of unbounded branching.}
We proved the lower bound for a proof system for $\IPC_\to$, but it can be generalized to analogous proof systems for
some stronger logics, namely implicational fragments of \emph{superintuitionistic (si) logics of unbounded branching}.
A si logic~$L$ has \emph{branching at most~$k$} if it is complete w.r.t.\ a class of finite Kripke models such that every
node has at most~$k$ immediate successors (or if it is included in such a logic); if $L$ does not have branching at
most~$k$ for any~$k\in\omega$, it has \emph{unbounded branching}. We consider $\nmi$ extended with finitely many axiom
schemata as proof systems for such logics. Any implicational logic of unbounded branching is included in $\lgc{BD}_2$
(the logic of Kripke models of depth~$2$), which can be axiomatized over $\IPC$ by the schema
\begin{equation}\label{eq:7}
((\fii\to((\psi\to\chi)\to\psi)\to\psi)\to\fii)\to\fii
\end{equation}
(this is an implicational version of the more familiar axiom $\fii\lor(\fii\to(\psi\lor\neg\psi))$). It is not a priori
clear that the implicational fragment of $\lgc{BD}_2$ is also axiomatized by \eqref{eq:7} over~$\IPC_\to$, but this can
be shown using the criterion in \cite[L.~4.11]{ej:implic}. Thus, it suffices to prove our lower bound for $\nmi$
extended with axioms~\eqref{eq:7}. This can be done by a minor modification of the proof of Theorem~\ref{thm:fmi-imp}: for
each instance $\omega$ of~\eqref{eq:7} used in~$\Pi$, we include $\omega$ itself as well as
$\fii\to((\psi\to\chi)\to\psi)\to\psi$ in~$P$. These formulas are classically valid, hence they will not affect the
final argument showing that $C$ interpolates $\neg\alpha_1(\neg\vec p,\vec r)\to\alpha_0(\vec p,\vec q)$, and their
presence in~$P$ easily implies $\Sls_{I,J}\omega$.

\paragraph{Full propositional language.}
It is straightforward to generalize dag-like natural deduction to the full language $\{\to,\land,\lor,\bot\}$ of
intuitionistic logic, including a suitable version of Lemma~\ref{lem:A}. The lower bound still holds for this proof
system: we can extend Definition~\ref{def:sls} using the standard Kleene slash conditions
\begin{align*}
\sls(\fii\land\psi)&\iff\sls\fii\text{ and }\sls\psi,\\
\sls(\fii\lor\psi)&\iff\Sls\fii\text{ or }\Sls\psi,
\end{align*}
and $\nsls\bot$; then we can prove the analogue of Lemma~\ref{lem:sls-sound}, and the rest of the
argument goes through unchanged.

The only problem is that this generalization interferes with the extension to logics of unbounded branching from the
previous paragraph. While positive fragments (i.e., $\{\to,\land,\lor\}$) of logics of unbounded branching are still
included in $\lgc{BD}_2$, this is not true for fragments including~$\bot$: then we only get that logics of unbounded
branching are included in either $\lgc{BD}_2$ or $\lgc{KC}+\lgc{BD}_3$ (see \cite[Thm.~6.9]{ej:sfef}; $\lgc{KC}$
denotes the logic of weak excluded middle). The proof of the
lower bound in the full language works fine for logics included in~$\lgc{BD}_2$ as indicated above, but unfortunately
we do not know a direct way of proving it for $\lgc{KC}+\lgc{BD}_3$. It seems that in this case we need the reduction
to the $\bot$-free fragment as given in \cite[L.~6.30]{ej:sfef} or \cite[\S4.1]{ej:implic}.

\paragraph{Frege and Extended Frege.}
As we already mentioned in the introduction, the result applies to the Frege system for~$\IPC_\to$, as this is
essentially a fragment of $\nmi$ without the $\imi$ rule (see Theorem~\ref{thm:sim-f-nm} in the appendix for more details).
However, the argument can be adapted to Frege systems directly, using closure under modus ponens \eqref{eq:mp} in place
of $\Pi$-closure. This also works for Frege systems of si logics included in~$\lgc{BD}_2$ in the full propositional
language as explained above. Since the lower bound is on the number of lines rather than overall proof size, it also
applies to Extended Frege systems.

\paragraph{Separation from Substitution Frege.}
We have only shown that the $\tau_n$ formulas are $\IPC_\to$ tautologies, but more constructively, they have polynomial-size (and
polynomial-time constructible) proofs in the \emph{Substitution Frege} proof system for~$\IPC_\to$. This can be
demonstrated along the lines of the proof of \cite[Thm.~4.22]{ej:implic} or \cite[L.~6.29]{ej:sfef}. Thus, for all
proof systems subject to the lower bound, we actually obtain an exponential separation from the $\IPC_\to$ Substitution
Frege system.

\paragraph{Larger bounds.}
The Colouring--Cocolouring tautologies can be made shorter using bit encoding of the colouring functions:
instead of the variables $q_{il}$ for $i<n$, $l<k$ as in Definition~\ref{def:taut}, we use variables $q_{ile}$ for
$i<n$, $l<\cl{\log k}$, and $e\in\{0,1\}$, with intended meaning ``the $l$th bit of the colour assigned to node $i$ is
$e$'', and likewise for~$\vec r$. This reduces the size of~$\tau_n$ to $O(n^2\log n)$ while keeping the same
proof size lower bound in terms of~$n$, thus the lower bound in terms of $\lh\fii$ improves to
$2^{\lh\fii^{1/16-o(1)}}$ (i.e., $\Omega(2^{\lh\fii^{1/16-\ep}})$ for arbitrarily small $\ep>0$).

Instead of Colouring--Cocolouring tautologies, we can use tautologies based on the original Clique--Colouring disjoint
pair as in \cite{hru:lbmod,hru:lbint,hru:nonclas,ej:sfef,ej:implic} (and in a preliminary version of this paper). They
have larger size, viz.\ $O(n^2k^2)$, but the monotone circuit size lower bound increases even more to $2^{\Omega(k^{1/2})}$
by Alon--Boppana~\cite{alonbop}. For $k\approx\sqrt n$, this improves the bound in Theorem~\ref{thm:main} to
$2^{\Omega(\lh\fii^{1/12})}$; if we raise $k$ to $\approx n^{2/3-o(1)}$ 
(the largest value to which the Alon--Boppana result applies), it improves further to $2^{\lh\fii^{1/10-o(1)}}$.

Any improvements of the underlying monotone circuit size lower bounds directly translate to improvements of the proof
size lower bounds. Recently, de Rezende and Vinyals~\cite{rez-vin:col-sun} proved a strengthening of the Alon--Boppana
lower bound to $n^{\Omega(k)}$ for $k\le n^{1/2-o(1)}$, and of the Hrube\v s--Pudl\'ak bound (our
Theorem~\ref{thm:hru-pud}) to $2^{k^{1/2-o(1)}}$. This implies improvements of Theorem~\ref{thm:main} to
$2^{\lh\fii^{1/10-o(1)}}$ for the Colouring--Cocolouring tautologies from Definition~\ref{def:taut},
$2^{\lh\fii^{1/8-o(1)}}$ for the bit-encoded Colouring--Cocolouring tautologies, and $2^{\lh\fii^{1/6-o(1)}}$ for
Clique--Colouring tautologies with $k=n^{1/2-o(1)}$. In fact, their results apply to a restricted version of the
Clique--Colouring problem where the graph of size $n=(k+1)m$ is $(k+1)$-partite with each element of the clique chosen
from a specific part of size $m\approx k^{1+o(1)}$, and the colours of nodes from a given part are chosen from a
palette of constant size; moreover, each colour occurs in the palettes of only $O(1)$ parts. The corresponding
$\IPC_\to$ tautologies have size $O(km^2)=O(k^{3+o(1)})$, yielding a $2^{\lh\fii^{1/3-o(1)}}$ proof size lower bound.

The best circuit size lower bounds one could hope to achieve with this line of reasoning would be a $2^{\Omega(k)}$
bound on Clique--Colouring with $k$ a constant fraction of~$n$ (i.e., with $m=O(1)$), implying a $2^{\Omega(k)}$
bound on Colouring--Cocolouring. These would translate to a $2^{\Omega(\lh\fii^{1/5})}$ proof size lower bound for the
Colouring--Cocolouring tautologies, $2^{\Omega(\lh\fii^{1/4-o(1)})}$ for bit-encoded Colouring--Cocolouring,
$2^{\Omega(\lh\fii^{1/4})}$ for Clique--Colouring with $k=\Theta(n)$, and an optimal $2^{\Omega(\lh\fii)}$ lower bound
for the restricted Clique--Colouring tautologies, matching the basic $2^{O(\lh\fii)}$ upper bound on the size of
intuitionistic proofs. (All these bounds are essentially tight for the respective tautologies.)

\section{Conclusion}\label{sec:conclusion}

We have shown how to prove a disjunction-free formulation of feasible disjunction property for implicational
intuitionistic logic directly using an efficient version of Kleene's slash, without reintroducing disjunctions into the
proof. More generally, we demonstrated an implicational version of Hrube\v s-style feasible monotone interpolation, and
exploited it to prove exponential lower bounds on the number of lines in dag-like natural deduction $\nmi$ for
intuitionistic implicational logic (or equivalent familiar systems such as Frege). This provides a simple refutation of
Gordeev and Haeusler's claims that all $\IPC_\to$ tautologies have polynomial-size proofs in~$\nmi$ that should be
accessible to a broad logic-aware audience.

Our approach consolidated the proof-theoretic components of the exponential lower bound to a single argument, obviating
the need for translation of intuitionistic logic to its implicational fragment, or of dag-like natural deduction to
Frege systems. The lower bound is not fully self-contained as we still rely on monotone circuit lower bounds;
this combinatorial component of our lower bound has a quite different flavour from the proof-theoretic part and uses
quite different techniques, thus it does not look very promising to try to combine them. Fortunately, we believe there
is no pressing need for that, as monotone circuit bounds are now a fairly well-understood part of standard literature.
The proof of the original Alon--Boppana bound in~\cite{alonbop} is neither long nor difficult to follow;
likewise, the relevant arguments in \cite{hru-pud:col-col,juk:boolf} are easily accessible.

\section*{Acknowledgments}
I want to thank Pavel Hrube\v s for the suggestion to use tautologies based on the Colouring--Cocolouring principle,
Susanna de Rezende for kindly explaining her work, and the anonymous referee for improvements in the presentation.

The research was supported by the Czech Academy of Sciences (RVO 67985840) and GA \v CR project 23-04825S.

\bibliographystyle{mybib}
\bibliography{impsimp}

\appendix

\section{Equivalence with Frege}\label{sec:equiv-with-frege}

Our objective in Section~\ref{sec:lower-bound} was to prove an exponential lower bound on the size of $\nmi$-proofs as
directly as we could, and in particular, we avoided translation of $\nmi$ to other proof systems such as Frege.
However, no treatment of the proof complexity of $\nmi$ can be complete without showing that it is, after all,
polynomially equivalent to the (intuitionistic implicational) Frege proof system~$\fri$. This is implicit in
Reckhow~\cite{reck} and Cook and Reckhow~\cite{cookrek}, but they work with a different formulation of natural
deduction, and with classical logic, hence it is worthwhile to spell out the reduction adapted to our situation, which
is the main goal of this section (Theorems \ref{thm:sim-f-nm}, \ref{thm:sim-nm-f}, and~\ref{thm:sim-nm-f-conj}).

Let us mention that even though we formulate the results in this and the next section as only bounds on proof size (and
other parameters), they are all constructive in that the relevant proofs can be computed by polynomial-time algorithms.

We start by defining the intuitionistic implicational Frege system~$\fri$.
\begin{Def}\label{def:frege}
A \emph{(sequence-like) $\fri$-derivation} of $\fii\in\Form$ from $\Gamma\sset\Form$ is a finite sequence of formulas
$\Pi=\p{\gamma_i:i<t}$ such that $t>0$, $\gamma_{t-1}=\fii$, and for each $i<t$: $\gamma_i\in\Gamma$, or $\gamma_i$ is an
instance of one of the logical axioms
\begin{align}
\tag{A1}\label{eq:ax1}\alpha&\to\beta\to\alpha,\\
\tag{A2}\label{eq:ax2}(\alpha\to\beta\to\gamma)&\to(\alpha\to\beta)\to(\alpha\to\gamma)
\end{align}
for some $\alpha,\beta,\gamma\in\Form$, or $\gamma_i$ is derived from $\gamma_j$ and~$\gamma_k$ for some $j,k<i$ by the
rule of modus ponens
\begin{equation}
\tag{MP}\label{eq:mp}\alpha,\alpha\to\beta\ru\beta,
\end{equation}
i.e., $\gamma_k=(\gamma_j\to\gamma_i)$. The \emph{number of lines} of~$\Pi$ is~$t$, and the \emph{size} of~$\Pi$ is
$\dlh\Pi=\sum_{i<t}\lh{\gamma_i}$. 

A \emph{dag-like $\fri$-derivation} of $\fii$ from~$\Gamma$ is $\Pi=\p{V,E,\gamma}$, where $\p{V,E}$ is a finite dag
with a unique node $\roo$ of out-degree~$0$ (the \emph{root}), all nodes have in-degree $0$ (the \emph{axioms} or
\emph{leaves}) or~$2$ (the \emph{\eqref{eq:mp}-nodes}), $\gamma=\p{\gamma_v:v\in V}$ is a labelling of nodes by
formulas such that $\gamma_\roo=\fii$, all leaves are labelled with elements of~$\Gamma$ or instances of \eqref{eq:ax1}
or~\eqref{eq:ax2}, and if $v$ is an \eqref{eq:mp}-node with premises $v_0$ and~$v_1$, then $\gamma_v$ is derived from
$\gamma_{v_0}$ and $\gamma_{v_1}$ by \eqref{eq:mp}. The \emph{number of lines} of~$\Pi$ is~$\lh V$, and the \emph{size}
of~$\Pi$ is $\dlh\Pi=\sum_{v\in V}\lh{\gamma_v}$.

A sequence-like or dag-like \emph{$\fri$-proof of~$\fii$} is a sequence-like or dag-like (resp.) $\fri$-derivation of
$\fii$ from~$\nul$.

The \emph{height} of a dag-like $\fri$-derivation or $\nmi$-derivation $\p{V,E,\gamma}$ is the maximal length of a
directed path from a leaf to the root. Such a derivation is \emph{tree-like} if the underlying dag $\p{V,E}$ is a tree,
i.e., all nodes have out-degree at most~$1$. Tree-like $\fri$-derivations and $\nmi$-derivations are also called
\emph{$\fri^*$-derivations} and \emph{$\nmi^*$-derivations} (respectively), and likewise for $\fri^*$-proofs and
$\nmi^*$-proofs.

The \emph{formula size} of a dag-like $\fri$-derivation or $\nmi$-derivation $\p{V,E,\gamma}$ is $\max_{v\in
V}\lh{\fii_v}$, and likewise for sequence-like $\fri$-derivations.
\end{Def}

Observe that $\nmi^*$ is the implicational fragment of the standard natural deduction system. It is well known that
sequence-like and dag-like Frege are just different presentations of the same proof system:
\begin{Lem}\label{lem:f-seq-dag}
A sequence-like (dag-like) $\fri$-derivation of $\fii$ from~$\Gamma$ can be converted to a dag-like (sequence-like,
resp.) $\fri$-derivation of $\fii$ from~$\Gamma$ with at most the same size, number of lines, and formula size. 
\end{Lem}
\begin{Pf}
Given a sequence-like derivation $\p{\gamma_i:i<t}$ of $\fii$ from~$\Gamma$, put $V=[t]$. Let $I$ be the set of $i<t$
such that $\gamma_i$ is not an axiom (from $\Gamma$, or an instance of \eqref{eq:ax1} or~\eqref{eq:ax2}); for each
$i\in I$, fix $i_0,i_1<i$ such that $\gamma_i$ is derived from $\gamma_{i_0}$ and $\gamma_{i_1}$ by~\eqref{eq:mp}, and
let $E=\{\p{i_j,i}:i\in I,j\in\{0,1\}\}$. Observe that $\p{V,E}$ is acyclic as $E\sset{<}\res[t]$. Then
$\p{V,E,\p{\gamma_i:i<t}}$ is a dag-like $\fri$-derivation of $\fii$ from~$\Gamma$, possibly after eliminating nodes
from which the root $t-1$ is not reachable.

Conversely, let $\p{V,E,\gamma}$ be a dag-like $\fri$-derivation of $\fii$ from~$\Gamma$, and $t=\lh V$. Since
$\p{V,E}$ is acyclic, we can find an enumeration $V=\{v_i:i<t\}$ such that $E\sset\{\p{v_i,v_j}:i<j\}$ (a ``topological
ordering'' of $\p{V,E}$). The root $\roo\in V$ is the only node without a successor, hence we must have $\roo=v_{t-1}$.
Then $\p{\gamma_{v_i}:i<t}$ is a sequence-like $\fri$-derivation of $\fii$ from~$\Gamma$.
\end{Pf}

The sequence-like definition is simpler, and is usually taken as the official definition of Frege systems (we follow
this usage). Nevertheless, the dag-like definition has other benefits, in particular it allows the introduction of
tree-like proofs and the height measure: this cannot be done directly with sequence-like proofs as it depends on the
choice of the dag structure, which may not be uniquely determined by the proof sequence alone.

Let us also note basic dependencies between the various proof parameters:
\begin{Obs}\label{obs:param}
An $\nmi$- or (dag-like) $\fri$-derivation with formula size~$r$ and $t$~lines has size at most~$rt$ and at least
$\max\{r,t\}$. A derivation with height~$h$ has less than $2^{h+1}$ lines.
\end{Obs}
\begin{Pf}
The first part is obvious. In a dag with in-degree~$2$ and root~$\roo$, there are at most $2^l$ paths of length~$l$
ending in~$\roo$. Thus, if $\roo$ is reachable from any node in at most $h$~steps, there are at most
$\sum_{l\le h}2^l=2^{h+1}-1$ nodes.
\end{Pf}

We mostly consider formula size to be an auxiliary measure that can be used to conveniently bound size as per
Observation~\ref{obs:param}; it is not that interesting on its own.

A simple, yet very useful, property of Frege and natural deduction systems is that instances of any derivable
schema have linear-size proofs. This is convenient for construction of asymptotically short proofs without worrying too
much about the choice of basic axioms: we can use \emph{any} valid schematic axioms and rules in a given argument as
long as the number of different schemata is kept fixed.
\begin{Lem}\label{lem:sch}
Fix $\Gamma\sset\Form$ and $\fii\in\Form$ in variables $\{p_i:i<k\}$ such that $\Gamma\vdash\fii$. Then for all
substitutions~$\sigma$, there are $\fri^*$-derivations and $\nmi^*$-derivations of $\sigma(\fii)$ from $\sigma(\Gamma)$
with $O(1)$ lines and size $O(s)$, where $s=\sum_{i<k}\lh{\sigma(p_i)}$. (The constants implied in the $O(\dots)$
notation depend on $\Gamma$ and~$\fii$.) Moreover, we may assume the derivations use each axiom from $\sigma(\Gamma)$
only once.
\end{Lem}
\begin{Pf}
Let $\Pi=\p{\gamma_i:i<t}$ be a fixed $\fri^*$-derivation of $\fii$ from~$\Gamma$ such that all variables occurring
in~$\Pi$ are among $\{p_i:i<k\}$. Then for any substitution~$\sigma$, $\p{\sigma(\gamma_i):i<t}$ is an $\fri$-derivation
of $\sigma(\fii)$ from $\sigma(\Gamma)$ with $t$~lines and size at most $\Dlh\Pi s$. The argument for~$\nmi^*$ is
completely analogous.

Instead of applying the argument directly to $\Gamma\vdash\fii$, we may apply it to the $\IPC_\to$ tautology
$\vdash\Gamma\to\fii$. This yields tree-like proofs of $\sigma(\Gamma\to\fii)$ with $O(1)$~lines and size $O(s)$,
which we can turn into derivations of $\sigma(\fii)$ from~$\sigma(\Gamma)$ by $\lh\Gamma$ applications of
\eqref{eq:mp}/$\ime$; this ensures that each axiom from $\sigma(\Gamma)$ is used only once.
\end{Pf}

The simulation of $\fri$ by~$\nmi$ is completely straightforward:
\begin{Thm}\label{thm:sim-f-nm}
If $\fii$ has a dag-like $\fri$-derivation from~$\Gamma$ with $t$~lines, height~$h$, formula size~$r$, and size~$s$,
then $\fii$ has an $\nmi$-derivation from~$\Gamma$ with $O(t)$~lines, height~$h+O(1)$, formula size $O(r)$, and
size~$O(s)$. If the original $\fri$-derivation is tree-like, the $\nmi$-derivation can also be taken tree-like.
\end{Thm}
\begin{Pf}
Let $\Pi=\p{V,E,\gamma}$ be a dag-like $\fri$-derivation of $\fii$ from~$\Gamma$. Reinterpreting the
\eqref{eq:mp}-nodes as $\ime$-nodes, $\Pi$ becomes an $\nmi$-derivation from $\Gamma$ plus the instances of
\eqref{eq:ax1} and~\eqref{eq:ax2} that appear in~$\Pi$. By Lemma~\ref{lem:sch}, each of the latter can be
replaced by a tree-like $\nmi$-subproof with $O(1)$ lines (thus height $O(1)$) and size linear in the size of the axiom
instance, yielding an $\nmi$-derivation of $\fii$ from~$\Gamma$ with the stated parameters.
\end{Pf}

For the converse simulation of $\nmi$ by~$\fri$, we will need proofs of some auxiliary formulas. As proved in
\cite[L.~2.3]{ej:implic}, there are short proofs of ``structural rules'' for $\Gamma\to\fii$, showing in particular
that we can arbitrarily reorder $\Gamma$ so that we can treat it as a set. We include here optimized proofs of some
special cases.
\begin{Def}\label{def:subset-imp}
We extend the $\Gamma\to\fii$ notation to sequences indexed by finite subsets of integers. If $I\sset[m]$ and
$\Gamma=\p{\alpha_i:i\in I}=\p{\alpha_i}_{i\in I}$, we define $\Gamma\to\fii$ by induction on $\lh I$:
$\p{\alpha_i}_{i\in\nul}\to\fii$ is $\fii$, and if $I\ne\nul$, then $\p{\alpha_i}_{i\in I}\to\fii$ is
$\alpha_h\to\p{\alpha_i}_{i\in I\bez\{h\}}\to\fii$, where $h=\max I$. (I.e., $\p{\alpha_i}_{i\in I}\to\fii$
is $\p{\alpha_{i_j}}_{j<n}\to\fii$, where $\p{i_j:j<n}$ is an increasing enumeration of~$I$.)

If $\Gamma=\p{\alpha_i}_{i\in I}$, we put $\dom(\Gamma)=I$, $\lh\Gamma=\lh I$, and
$\dlh\Gamma=\sum_{i\in I}\lh{\alpha_i}$. We write $\Gamma\res J=\p{\alpha_i}_{i\in I\cap J}$. If
$\Delta=\p{\beta_i}_{i\in J }$, we write $\Gamma\sset\Delta$ when $I\sset J$ and $\alpha_i=\beta_i$ for all $i\in I$.
\end{Def}

First, a general observation that we will keep using to construct proofs of small height:
\begin{Lem}\label{lem:bal-tree}
Given a sequence of formulas $\p{\fii_i:i\le n}$, $n\ge1$, there is an $\fri^*$-derivation of $\fii_0\to\fii_n$ from
$\{\fii_i\to\fii_{i+1}:i<n\}$ with $O(n)$ lines, height $O(\log n)$, formula size $O(r)$, and size $O(rn)$ that uses
each assumption $\fii_i\to\fii_{i+1}$ only once, where $r=\max_i\lh{\fii_i}$.
\end{Lem}
\begin{Pf}
We arrange the implications in a balanced binary tree with $n$~leaves. Formally, we construct for each%
\footnote{In this paper, $\log$ denotes base-$2$ logarithm.}
$k\le\cl{\log n}$ and $i<n$ such that $2^k\mid i$ a derivation $\Pi^k_i$ of $\fii_i\to\fii_{\min\{i+2^k,n\}}$ by
induction on~$k$ as follows: $\Pi^0_i$ is the trivial derivation of $\fii_i\to\fii_{i+1}$ from itself. Let
$k<\cl{\log n}$ and $i<n$ be such that $2^{k+1}\mid i$. If $i+2^k\ge n$, we put $\Pi^{k+1}_i=\Pi^k_i$; otherwise, we
combine $\Pi^k_i$ and $\Pi^k_{i+2^k}$ to $\Pi^{k+1}_i$ using an instance of the schematic rule
\[\alpha\to\beta,\beta\to\gamma\vdash\alpha\to\beta,\]
i.e., an $\fri^*$-derivation of $\fii_i\to\fii_{\min\{i+2^{k+1},n\}}$ from $\fii_i\to\fii_{i+2^k}$ and
$\fii_{i+2^k}\to\fii_{\min\{i+2^{k+1},n\}}$ with $O(1)$ lines and size $O(r)$ that uses each assumption only once,
which exists by Lemma~\ref{lem:sch}.

Then $\Pi^{\cl{\log n}}_0$ is the desired derivation of $\fii_0\to\fii_n$.
\end{Pf}
\begin{Lem}\label{lem:subset}
Given sequences of formulas $\Gamma$ and~$\Delta$ such that $\Delta\sset\Gamma$, and $\fii\in\Form$, there exists an
$\fri^*$-proof of
\begin{equation}
\label{eq:8} (\Delta\to\fii)\to(\Gamma\to\fii)
\end{equation}
with $O(n)$ lines, height $O(\log n)$, formula size $O(s)$, and size $O(sn)$, where $n=\max\bigl\{\lh\Gamma,2\bigr\}$
and $s=\dlh\Gamma+\lh\fii$.
\end{Lem}
\begin{Pf}
We may assume $\Gamma=\p{\alpha_i}_{i<n}$ and $\Delta=\p{\alpha_i}_{i\in I}$, $I\sset[n]$. For each $i\le n$, let
$\fii_i$ denote the formula $(\Delta\res[i]\to\fii)\to(\Gamma\res[i]\to\fii)$. Then $\fii_i\to\fii_{i+1}$ is an
instance of one of the schemata
\begin{align*}
(\delta\to\gamma)&\to(\delta\to\alpha\to\gamma),\\
(\delta\to\gamma)&\to((\alpha\to\delta)\to\alpha\to\gamma)
\end{align*}
with $\delta=(\Delta\res[i]\to\fii)$, $\gamma=(\Gamma\res[i]\to\fii)$, and $\alpha=\alpha_i$, depending on whether
$i\in I$. Thus, it has an $\fri^*$-proof with $O(1)$ lines and size $O(s)$ by Lemma~\ref{lem:sch}. Using
Lemma~\ref{lem:bal-tree}, we can combine these proofs to a proof of $\fii_0\to\fii_n$ with $O(n)$ lines, height
$O(\log n)$, formula size $O(s)$, and size $O(sn)$. Since $\fii_n$ is~\eqref{eq:8}, it remains to detach the $\IPC_\to$
tautology
$\fii_0=(\fii\to\fii)$, which has a proof with $O(1)$ lines and size $O(\lh\fii)$.
\end{Pf}
\begin{Lem}\label{lem:set-weak}
Given sequences of formulas $\Gamma$, $\Delta$, and~$\Theta$, and $\fii,\psi\in\Form$, there are $\fri^*$-proofs of
\begin{align}
\label{eq:9} (\Gamma\to\fii\to\psi)\to(\Gamma\to\fii)&\to(\Gamma\to\psi),\\
\label{eq:10} \Gamma&\to(\Gamma\to\fii)\to\fii,\\
\label{eq:11} (\Gamma\to\Gamma\to\fii)&\to(\Gamma\to\fii),\\
\label{eq:15} (\Theta\to\Gamma\to\Delta\to\fii)&\to(\Theta\to\Delta\to\Gamma\to\fii)
\end{align}
with $O(n)$ lines, height $O(\log n)$, formula size $O(s)$, and size $O(sn)$, where
$n=\max\bigl\{\lh\Gamma+\lh\Delta+\lh\Theta,2\}$ and $s=\dlh\Gamma+\dlh\Delta+\dlh\Theta+\lh\fii+\lh\psi$.
\end{Lem}
\begin{Pf}
We prove \eqref{eq:9} using the same strategy as in Lemma~\ref{lem:subset}: putting
\[\fii_i=(\Gamma\res[i]\to\fii\to\psi)\to(\Gamma\res[i]\to\fii)\to(\Gamma\res[i]\to\psi)\]
for each $i\le\lh\Gamma$, $\fii_i\to\fii_{i+1}$ has a proof with $O(1)$ lines and size $O(s)$ as it is an instance of
the schema
\[(\beta\to\gamma\to\delta)\to((\alpha\to\beta)\to(\alpha\to\gamma)\to(\alpha\to\delta)).\]
These proofs combine to a proof of $\fii_0\to\fii_{\lh\Gamma}$ with the stated parameters using Lemma~\ref{lem:bal-tree}.
Then $\fii_{\lh\Gamma}$ is~\eqref{eq:9}, and $\fii_0=(\fii\to\psi)\to(\fii\to\psi)$ has a short proof.

For~\eqref{eq:10}, we put $\fii_i=((\Gamma\to\fii)\to\Gamma\res[i]\to\fii)\to\Gamma\res[i]\to(\Gamma\to\fii)\to\fii$.
Then $\fii_0$ is an instance of $\alpha\to\alpha$, and $\fii_i\to\fii_{i+1}$ is an instance of
\[((\alpha\to\beta)\to\delta)\to(\alpha\to\gamma\to\beta)\to\gamma\to\delta\]
(with $\alpha=(\Gamma\to\fii)$, $\beta=(\Gamma\res[i]\to\fii)$, $\gamma=\gamma_i$, and
$\delta=(\Gamma\res[i]\to(\Gamma\to\fii)\to\fii)$, where $\Gamma=\p{\gamma_i}_{i<\lh\Gamma}$).
Thus, using Lemmas \ref{lem:sch} and~\ref{lem:bal-tree}, we obtain an $\fri^*$-proof of $\fii_{\lh\Gamma}$ with $O(n)$ lines,
height $O(\log n)$, formula size $O(s)$, and size $O(sn)$. Detaching the premise $(\Gamma\to\fii)\to\Gamma\to\fii$
of~$\fii_{\lh\Gamma}$ yields~\eqref{eq:10}.

\eqref{eq:11} follows by \eqref{eq:mp} from \eqref{eq:10} and~\eqref{eq:9}.

\eqref{eq:15}: We have $(\Gamma\to\Delta\to\fii)\to(\Delta\to\Gamma\to\Delta\to\Gamma\to\fii)$ from~\eqref{eq:8}, and
$(\Delta\to\Gamma\to\Delta\to\Gamma\to\fii)\to(\Delta\to\Gamma\to\fii)$ from~\eqref{eq:11}, thus we obtain
\eqref{eq:15} when $\Theta=\nul$. The general case follows by applying~\eqref{eq:9}.
\end{Pf}
\begin{Thm}\label{thm:sim-nm-f}
If $\fii$ has an $\nmi$-derivation from~$\Gamma$ with $t$~lines, height~$h$, and size~$s$, then $\fii$ has a dag-like
$\fri$-derivation from~$\Gamma$ with $O(t^2)$~lines, height~$O(h)$, formula size~$O(s)$, and size $O(st^2)$. If the
original $\nmi$-derivation is tree-like, the $\fri$-derivation can be taken tree-like as well.
\end{Thm}
\begin{Pf}
Let $\Pi=\p{V,E,\gamma}$ be an $\nmi$-derivation of $\fii$ from~$\Gamma$. Let $\p{\gamma'_i}_{i<t'}$,
$t'\le t$, be an injective enumeration of the set $\{\gamma_v:v\in V\}$, and for each $v\in V$, let $A'_v$ denote
the sequence $\p{\gamma'_i:i<t',\gamma'_i\in A_v\bez\Gamma}$; notice that $\dlh{A'_v}\le s$.
We consider the collection of $\IPC_\to$ tautologies $\p{A'_v\to\gamma_v:v\in V}$, and complete it to a valid $\fri$-derivation
from~$\Gamma$ using Lemmas \ref{lem:subset} and~\ref{lem:set-weak}.

In more detail, for every $v\in V$, we construct an $\fri^*$-derivation $\Pi_v$ of $A'_v\to\gamma_v$ from
$\{A'_u\to\gamma_u:\p{u,v}\in E\}\cup\Gamma$ with $O(t)$~lines, height $O(\log t)$, and formula size $O(s)$. Moreover,
each assumption $A'_u\to\gamma_u$ is used only once, and the path from it to the conclusion has length $O(1)$; both of
these properties are obtained by constructing a derivation of $\p{A'_u\to\gamma_u}_{\p{u,v}\in E}\to A'_v\to\gamma_v$
from~$\Gamma$ and applying~\eqref{eq:mp}:
\begin{itemize}
\item If $v$ is a leaf, then either $\gamma_v\in\Gamma$ and $A'_v=\nul$, in which case we take the trivial derivation
of $\gamma_v$ from itself, or $A'_v=\p{\gamma_v}$, in which case we find an $\fri^*$-proof of $\gamma_v\to\gamma_v$
with $O(1)$ lines and size $O(s)$ by Lemma~\ref{lem:sch}.
\item If $v$ is an $\imi$-node with premise~$u$, we have $\gamma_v=(\alpha\to\beta)$ and $\gamma_u=\beta$ for some
$\alpha$ and~$\beta$, and $A'_v=A'_u\bez\{\alpha\}$ as a set. If $\alpha\in A'_u$, then
$(A'_u\to\beta)\to(A'_v\to\alpha\to\beta)$ is an instance of~\eqref{eq:15}, otherwise it is an instance
of~\eqref{eq:8}.
\item If $v$ is an $\ime$-node with premises $u_0$ and~$u_1$, then $\gamma_{u_0}=\alpha$,
$\gamma_{u_1}=(\alpha\to\beta)$, and $\gamma_v=\beta$ for some $\alpha$ and~$\beta$. We have $A'_{u_i}\sset A'_v$,
hence \eqref{eq:8} gives $\fri^*$-proofs of $(A'_{u_0}\to\alpha)\to A'_v\to\alpha$ and $(A'_{u_1}\to\alpha\to\beta)\to
A'_v\to\alpha\to\beta$. We infer $(A'_{u_1}\to\alpha\to\beta)\to(A'_{u_0}\to\alpha)\to A'_v\to\beta$ using the instance
$(A'_v\to\alpha\to\beta)\to(A'_v\to\alpha)\to A'_v\to\beta$ of~\eqref{eq:9} and $O(1)$ additional proof lines by
Lemma~\ref{lem:sch}.
\end{itemize}
Combining these derivations~$\Pi_v$ along the shape of the original derivation~$\Pi$ yields an $\fri$-derivation
(tree-like if $\Pi$ is tree-like) of $\fii$ from~$\Gamma$ with $O(t^2)$ lines, height $O(h+\log t)=O(h)$ (cf.\
Observation~\ref{obs:param}), formula size $O(s)$, and size $O(st^2)$ as promised.
\end{Pf}

The bottleneck in the proof of Theorem~\ref{thm:sim-nm-f} is that formulas of the form $\Gamma\to\fii$ with long~$\Gamma$
are cumbersome to operate as $\fii$ is nested deep inside, and when untangling it we need to keep copying large parts
of the formula. This could be avoided if we had a conjunction connective: using $\ET\Gamma\to\fii$ instead,
$\fii$ sits right at nesting depth~$1$; if we arrange the big conjunction $\ET\Gamma$ in a balanced binary tree,
the individual entries of~$\Gamma$ are also easy to access at nesting depth $O(\log n)$, and wholesale
manipulations such as Lemma~\ref{lem:subset} can be done using a divide-and-conquer approach that saves size.

We do not have $\land$ in implicational logic, as it is not definable in terms of~$\to$. However, we may observe that
if we fix a formula~$\fii$, then formulas $\alpha,\beta$ of the form $\Phi\to\fii$ \emph{do} have a definable
conjunction operation: $\alpha$ is equivalent to $(\alpha\to\fii)\to\fii$, and likewise for~$\beta$, thus also
$\alpha\land\beta$ is equivalent to $(\alpha\land\beta\to\fii)\to\fii$, which can be written as
$(\alpha\to\beta\to\fii)\to\fii$. This idea was introduced in \cite[Prop.~2.6]{ej:implic} to prove polynomial
simulation of Frege by tree-like Frege for purely implicational logic (cf.\ Theorem~\ref{thm:fr-tree}), but here we
will use it to improve the bounds in Theorem~\ref{thm:sim-nm-f}.
\begin{Def}\label{def:rel-conj}
For any formulas $\fii$, $\alpha$, and~$\beta$, we put
\begin{align*}
\alpha^\fii&=(\alpha\to\fii)\to\fii,\\
\alpha\bret\fii\beta&=(\alpha\to\beta\to\fii)\to\fii.
\end{align*}
For all sequences of formulas $\Gamma=\p{\alpha_i:i\in I}$, $I\sset[m]$, we define $\RET\fii{i\in I}\alpha_i$, also
denoted $\RET\fii{}\Gamma$, by induction on~$m$:
\[\RET\fii{i\in I}\alpha_i=\begin{cases}
\top,&I=\nul,\\
\alpha_{i_0}^\fii,&I=\{i_0\},\\[\medskipamount]
\displaystyle\RET\fii{i\in I-2^k}\alpha_{2^k+i},&I\sset[2^k,2^{k+1}),\\[\bigskipamount]
\displaystyle\Bigl(\RET\fii{\!\!i\in I\cap[2^k]\!\!}\alpha_i\Bigr)\bret\fii\Bigl(\RET\fii{\!\!i\in I-2^k\!\!}\alpha_{2^k+i}\Bigr),
&I\sset[2^{k+1}],I\cap[2^k]\ne\nul\ne I\cap[2^k,2^{k+1}),
\end{cases}\]
where $\top$ is a fixed $\IPC_\to$ tautology, $k\ge0$, and $I-2^k=\{i:2^k+i\in I\}$. We write
$\RET\fii{i<n}\alpha_i$ for $\RET\fii{i\in[n]}\alpha_i$.
\end{Def}

The idea is that $\RET\fii{i<2^k}\alpha_i$ consists of $\bret\fii$ arranged in a perfect binary tree of height~$k$,
while if $I\sset[2^k]$, then $\RET\fii{i\in I}\alpha_i$ conforms to the same arrangement except that unused leaves and
non-splitting inner nodes are omitted; this ensures that the layouts of $\RET\fii{i\in I}\alpha_i$ and
$\RET\fii{i\in J}\alpha_i$ for any $I,J\sset[2^k]$ are compatible, facilitating efficient manipulation of
$\RET\fii{}\Gamma$ in a divide-and-conquer manner.
\begin{Lem}\label{lem:ret-size}
The size of $\RET\fii{}\Gamma$ is $\dlh\Gamma+O\bigl(\LH\fii n\bigr)$, where $n=\max\bigl\{\lh\Gamma,1\bigr\}$.
\end{Lem}
\begin{Pf}
Observe that the inductive definition introduces $\bret\fii$ only when the sequences on both sides are nonempty. Thus,
$\RET\fii{}\Gamma$ is a binary tree of $\bret\fii$ with $n$~leaves where every inner node splits, thus there are
$n-1$ inner nodes. Since $\alpha$ and~$\beta$ occur only once in~$\alpha\bret\fii\beta$, each node of the tree gives
rise to only one subformula of $\RET\fii{}\Gamma$; thus, $\RET\fii{}\Gamma$ consists of one occurrence of
each~$\alpha_i$ of total size $\dlh\Gamma$, and $O(1)$ occurrences of $\fii$ and~$\to$ per each node of the
tree of total size $O\bigl(\LH\fii n\bigr)$.
\end{Pf}

The following is a $\RET\fii{}$-version of Lemma~\ref{lem:subset} that also handles unions of two sequences.
\begin{Lem}\label{lem:subset-conj}
Let $\fii\in\Form$ and $\Gamma=\p{\alpha_i:i\in I}$ be a sequence of formulas with $\lh\Gamma=n\ge1$ and $I\sset[m]$,
$m\ge2$. Let $\Gamma_u=\Gamma\res I_u$ for $u=0,1,2$, where $I_u\sset I$ are such that $I_2\sset I_0\cup I_1$. Then
there is an $\fri^*$-proof of
\begin{equation}\label{eq:12}
\RET\fii{}\Gamma_0\to\RET\fii{}\Gamma_1\to\RET\fii{}\Gamma_2
\end{equation}
with $O(n)$ lines, height $O(\log m)$, formula size $O(s+\LH\fii n)$, and size $O\bigl((s+\LH\fii n)\log m\bigr)$,
where $s=\dlh\Gamma$.
\end{Lem}
\begin{Pf}
We construct the proofs by induction on $\cl{\log m}$. If $m=2$ or $n=1$, then \eqref{eq:12} has a proof with $O(1)$
lines and size $O(s+\lh\fii)$ by Lemma~\ref{lem:sch}. If $I\sset[2^k,2^{k+1})$ for some~$k$, we can just apply the
induction hypothesis (without changing the proof) to $\Gamma'=\p{\alpha_{2^k+i}:i\in I-2^k}$ and $\Gamma'_u=\Gamma'\res
(I_u-2^k)$, as $\RET\fii{}\Gamma_u=\RET\fii{}\Gamma'_u$.

Assume that $I\sset[2^{k+1}]$ and $I^0,I^1\ne\nul$, where $I^0=I\cap[2^k]$ and $I^1=I-2^k$. For each $u<3$, put
$I^0_u=I_u\cap[2^k]$ and $I^1_u=I_u-2^k$. Let $\Gamma^0=\Gamma\res I^0$, $\Gamma^1=\p{\alpha_{2^k+i}:i\in I^1}$, and
$\Gamma^v_u=\Gamma^v\res I^v_u$ for each $v<2$, $u<3$. There are proofs of
\begin{equation}\label{eq:13}
\RET\fii{}\Gamma_u\to\RET\fii{}\Gamma^v_u,\qquad\RET\fii{}\Gamma^0_u\to\RET\fii{}\Gamma^1_u\to\RET\fii{}\Gamma_u
\end{equation}
with $O(1)$ lines and size $O(s+\LH\fii n)$ using Lemma~\ref{lem:sch}: if $I^v_u=\nul$ for some $v<2$, then
$\RET\fii{}\Gamma_u=\RET\fii{}\Gamma^{1-v}_u$ and $\RET\fii{}\Gamma^v_u=\top$; otherwise, $\RET\fii{}\Gamma_u$ is
$\bigl(\RET\fii{}\Gamma^0_u\bigr)\bret\fii\bigl(\RET\fii{}\Gamma^1_u\bigr)$, thus \eqref{eq:13} are instances of the
valid schemata $(\alpha^0\to\fii)\bret\fii(\alpha^1\to\fii)\to(\alpha^v\to\fii)$ and
$\alpha\to\beta\to\alpha\bret\fii\beta$ (observe that each $\RET\fii{}\Gamma^v_u$ is of the form $\alpha\to\fii$ for
some formula~$\alpha$).

Using~\eqref{eq:13}, we can construct proofs of
\begin{equation}\label{eq:14}
\Bigl(\RET\fii{}\Gamma^0_0\to\RET\fii{}\Gamma^0_1\to\RET\fii{}\Gamma^0_2\Bigr)\to
\Bigl(\RET\fii{}\Gamma^1_0\to\RET\fii{}\Gamma^1_1\to\RET\fii{}\Gamma^1_2\Bigr)\to
\Bigl(\RET\fii{}\Gamma_0\to\RET\fii{}\Gamma_1\to\RET\fii{}\Gamma_2\Bigr)
\end{equation}
with $O(1)$ lines and size $O(s+\LH\fii n)$. The induction hypothesis for $\Gamma^0$ and~$\Gamma^1$ gives us proofs of
$\RET\fii{}\Gamma^v_0\to\RET\fii{}\Gamma^v_1\to\RET\fii{}\Gamma^v_2$ for $v<2$, and these together yield~\eqref{eq:12}.

We can imagine the resulting proof as a binary tree of \eqref{eq:14}~inferences. Since each application
of~\eqref{eq:14} corresponds to splitting $I$ to two nonempty disjoint subsets, each inner node has two children, and
the tree has at most $n$~leaves. Thus, the proof has $O(n)$ lines. Each application of~\eqref{eq:14} also strictly
decreases $\cl{\log m}$, hence the height of the proof is $O(\log m)$. The formula size is $O(s+\LH\fii n)$ using
Lemma~\ref{lem:ret-size}.

As for the size of the proof, the root of the tree contributes $O(s+\LH\fii n)$. Its two children contribute
$O(s_0+\LH\fii n_0)$ and $O(s_1+\LH\fii n_1)$, where $s_0+s_1=s$ and $n_0+n_1=n$, thus $O(s+\LH\fii n)$ together.
Continuing the same way, each level of the tree consists of inferences of size $O(s+\LH\fii n)$, and there are at most
$O(\log m)$ levels, hence the total size is $O\bigl((s+\LH\fii n)\log m\bigr)$. (More formally, we can prove such a
bound by induction on $\cl{\log m}$.)
\end{Pf}

We cannot use $\RET\fii{}A'_v\to\gamma_v^\fii$ with a fixed formula~$\fii$ instead of $A'_v\to\gamma_v$ for the
simulation of $\nmi$ by~$\fri$ as in the proof of Theorem~\ref{thm:sim-nm-f}, 
because the $\imi$-rule would translate to an unsound inference
\[\RET\fii{}A'_v\to\alpha^\fii\to\beta^\fii\vdash\RET\fii{}A'_v\to(\alpha\to\beta)^\fii.\]
We will in fact work with $\RET{\gamma_v}{}A'_v\to\gamma_v$, but this necessitates that we are able to transform
$\RET\fii{}\Gamma$ to $\RET\psi{}\Gamma$ for given $\fii$, $\psi$:
\begin{Lem}\label{lem:conj-trans}
Let $\fii,\psi\in\Form$ and $\Gamma=\p{\alpha_i:i\in I}$ be a sequence of formulas with $I\sset[m]$, $m\ge2$. Then
there is an $\fri^*$-proof of
\begin{equation}\label{eq:19}
\RET\fii{}\Gamma\to\Bigl(\RET\psi{}\Gamma\Bigr)^\fii
\end{equation}
with $O(n)$ lines, height $O(\log m)$, formula size $O(s+\LH\fii n+\LH\psi n)$, and size $O\bigl((s+\LH\fii n+\LH\psi
n)\log m\bigr)$, where $n=\max\bigl\{\lh\Gamma,1\bigr\}$ and $s=\dlh\Gamma$.
\end{Lem}
\begin{Pf}
We construct the proofs by induction on $\cl{\log m}$, similarly to Lemma~\ref{lem:subset-conj}. If $m=2$ or $n=1$, then
\eqref{eq:19} has a proof with $O(1)$ lines and size $O(s+\lh\fii+\lh\psi)$ by Lemma~\ref{lem:sch}. If $I\sset[2^k,2^{k+1})$ for
some~$k$, we can apply the induction hypothesis to $\Gamma'=\p{\alpha_{2^k+i}:i\in I-2^k}$ without changing the proof.
If $I\sset[2^{k+1}]$ and $I^0,I^1\ne\nul$, where $I^0=I\cap[2^k]$ and $I^1=I-2^k$, the induction hypothesis applied to
$\Gamma^0=\Gamma\res I^0$ and $\Gamma^1=\p{\alpha_{2^k+i}:i\in I^1}$ gives proofs of
\[\RET\fii{}\Gamma^0\to\Bigl(\RET\psi{}\Gamma^0\Bigr)^\fii,\qquad
  \RET\fii{}\Gamma^1\to\Bigl(\RET\psi{}\Gamma^1\Bigr)^\fii.\]
These yield~\eqref{eq:19} using an instance of the schema
\[(\alpha\to\beta^\fii)\to(\gamma\to\delta^\fii)\to\alpha\bret\fii\gamma\to(\beta\bret\psi\delta)^\fii\]
(we invite the reader to check this is indeed a valid schema).

The resulting proof has the stated size parameters by the same argument as in Lemma~\ref{lem:subset-conj}.
\end{Pf}

Before we get to the improved simulation of $\nmi$ by~$\fri$, we need to introduce one more size parameter so that we
can state the bounds accurately:
\begin{Def}\label{def:inf-size}
The \emph{inferential size} of a $\nmi$-derivation or dag-like $\fri$-derivation $\p{V,E,\gamma}$ is $\sum_{v\in
V}s_v$, where $s_v=\lh{\gamma_v}+\sum_{\p{u,v}\in E}\lh{\gamma_u}$.
\end{Def}

Clearly, a derivation with $t$~lines and formula size~$r$ has inferential size $O(rt)$. A tree-like derivation (or
more generally, a derivation where each node has bounded out-degree) of size~$s$ has inferential size $O(s)$. We will
see later (Lemma~\ref{lem:inf-size}) that any dag-like $\fri$-derivation of size~$s$ can be shortened to a derivation with
inferential size~$O(s)$, but we do not know whether the analogue for $\nmi$-derivations holds.
\begin{Thm}\label{thm:sim-nm-f-conj}
If $\fii$ has an $\nmi$-derivation from~$\Gamma$ with $t$~lines, height~$h$, formula size~$r$, and inferential
size~$\tilde s$, then $\fii$ has a dag-like $\fri$-derivation from~$\Gamma$ with $O(t^2)$~lines, height~$O(h)$, formula
size~$O(rt)$, and (inferential) size $O(\tilde st\log t)$. If the original $\nmi$-derivation is tree-like, the
$\fri$-derivation can be taken tree-like as well.
\end{Thm}
\begin{Pf}
We use the same notation and argument structure as in the proof of Theorem~\ref{thm:sim-nm-f}, but we work with the formulas
$\delta_v=\RET{\gamma_v}{}A'_v\to\gamma_v$ in place of $A'_v\to\gamma_v$. Observe
$\lh{\delta_v}=O\bigl(s+\LH{\gamma_v}t\bigr)=O(rt)$, where $s=\dlh\Pi$.

For each $v\in V$, we construct an $\fri^*$-derivation $\Pi_v$ of $\p{\delta_u}_{\p{u,v}\in E}\to\delta_v$ from~$\Gamma$
with $O(t)$ lines, height $O(\log t)$, formula size $O(s+s_vt)=O(rt)$, and size $O\bigl((s+s_vt)\log t\bigr)$, where
$s_v=\lh{\gamma_v}+\sum_{\p{u,v}\in E}\lh{\gamma_u}$:
\begin{itemize}
\item The case of $v$ being a leaf is straightforward.
\item If $v$ is an $\imi$-node with premise~$u$, we have $\gamma_v=(\alpha\to\beta)$ and $\gamma_u=\beta$ for some
$\alpha$ and~$\beta$, and $A'_u\sset A'_v\cup\{\alpha\}$ as a set. Lemma~\ref{lem:subset-conj} gives a proof of
$\RET\beta{}A'_v\to\alpha^\beta\to\RET\beta{}A'_u$, which (using $\alpha\to\alpha^\beta$) yields
$\delta_u\to\RET\beta{}A'_v\to\gamma_v$. Combining this with
$\RET{\gamma_v}{}A'_v\to\bigl(\RET\beta{}A'_v\bigr)^{\gamma_v}$ from Lemma~\ref{lem:conj-trans} gives
$\delta_u\to\RET{\gamma_v}{}A'_v\to\gamma_v$, i.e., $\delta_u\to\delta_v$.
\item If $v$ is an $\ime$-node with premises $u_0$ and~$u_1$, then $\gamma_{u_0}=\alpha$,
$\gamma_{u_1}=(\alpha\to\beta)$, and $\gamma_v=\beta$ for some $\alpha$ and~$\beta$, and $A'_{u_i}\sset A'_v$. Using
Lemmas \ref{lem:subset-conj} and~\ref{lem:conj-trans}, we obtain proofs of $\delta_{u_0}\to\RET\beta{}A'_v\to\alpha^\beta$ and
$\delta_{u_1}\to\RET\beta{}A'_v\to(\alpha\to\beta)^\beta$, which yield
$\delta_{u_0}\to\delta_{u_1}\to\RET\beta{}A'_v\to\beta$ (i.e., $\delta_{u_0}\to\delta_{u_1}\to\delta_v$) using the
schema $\alpha^\beta\to(\alpha\to\beta)^\beta\to\beta$.
\end{itemize}
Combining the $\Pi_v$ derivations yields an $\fri$-derivation (tree-like if $\Pi$ is tree-like) of $\fii$ from~$\Gamma$
with $O(t^2)$ lines, height $O(h)$, formula size $O(rt)$, and size $O\bigl(st\log t+\sum_vs_vt\log t)=O(\tilde st\log
t)$.
\end{Pf}
\begin{Rem}\label{rem:improve-nm}
We can improve the resulting $\fri$-derivation to a tree-like derivation of height $O(\log t)$ at the expense of
a mild size increase: see Theorem~\ref{thm:nm-tree}.

If we have a real~$\land$, the $\lh{\gamma_v}$ terms from the size parameters disappear, and we obtain a derivation
with formula size $O(s)$ and size $O(st\log t)$ rather than $O(\tilde st\log t)$. It is unclear how to achieve that
in the purely implicational setting. One possible improvement is to modify the inductive definition of $\RET\fii{}$ so
that $\RET\fii{}\Gamma=(\Gamma\to\fii)\to\fii$ whenever $\lh\Gamma\le\ell$, where $\ell\ge1$ is an extra parameter. Then
$\RET\fii{}\Gamma$ has size $O\bigl(\dlh\Gamma+\LH\fii\frac n\ell\bigr)$, where $n=\max\bigl\{\lh\Gamma,\ell\bigr\}$.
The proofs in Lemma~\ref{lem:subset-conj} will have formula size $O\bigl(s+\LH\fii\frac n\ell\bigr)$ and size
$O\bigl(s(\log m+\ell)+\LH\fii\frac n\ell\log m\bigr)$, and similarly for Lemma~\ref{lem:conj-trans}. In the context of
the proof of Theorem~\ref{thm:sim-nm-f-conj}, the optimal choice is $\ell\approx\sqrt{(\tilde s/s)\log t}$, which yields an
$\fri$-derivation of $\fii$ from~$\Gamma$ with $O(t^2)$ lines, height $O(\log t)$, formula size
$O\bigl(s+rt/\ell\bigr)$, and size $O\bigl(st\log t+\sqrt{\tilde sst^2\log t}\bigr)$.
\end{Rem}

\begin{Rem}\label{rem:ded-thm}
Using similar arguments, we can also prove an efficient version of Lemma~\ref{lem:ded}: if $\fii$ has an $\fri$-derivation
from $\Gamma=\{\alpha_i:i<n\}$ and~$\Delta$ with $t\ge n$ lines, height~$h$, formula size~$r$, and size~$s$, then
$\RET\fii{}\Gamma\to\fii$ and $\Gamma\to\fii$ have $\fri$-derivations from~$\Delta$ with $O(t)$~lines,
height~$O(h)$, formula size $O\bigl(r+\dlh\Gamma+\LH\fii n\bigr)$, and size $O\bigl(s+(\dlh\Gamma+\LH\fii n)t\bigr)$.
\end{Rem}

\section{Equivalence of dag-like and tree-like proofs}\label{sec:equivalence-dag-like}

Our final task is to show that $\nmi$ and~$\fri$ are polynomially equivalent to their tree-like versions $\nmi^*$ and
$\fri^*$; more precisely, we will show that an $\fri$-proof with $t$~lines can be converted to a polynomially larger
tree-like proof of height $O(\log t)$ (Theorem~\ref{thm:fr-tree}), which implies a similar simulation of $\nmi$ by $\nmi^*$
(Theorem~\ref{thm:nm-tree}).

The original argument by Kraj\'\i\v cek \cite[L.~4.4.8]{book} (stated in the context of classical logic, but
intuitionistically valid) relies on conjunctions: given a proof $\p{\gamma_i:i<t}$, we consider the conjunctions
$\tau_j=\ET_{i<j}\gamma_i$, construct short tree-like proofs of $\tau_i\to\tau_{i+1}$, and combine them to a proof
of $\tau_t$. A purely implicational version of the argument was sketched in \cite[Prop.~2.6]{ej:implic}, using the
$\alpha\bret\fii\beta$ formulas to emulate conjunctions. We now present the argument in detail, incorporating an
extra idea to save proof size: instead of (an implicational emulation of) the long conjunctions
$\tau_i\to\tau_{i+1}$, we start with $\tau'_i\to\gamma_i$ where $\tau'_i$ only consists of the premises needed to
infer~$\gamma_i$, and we gradually merge these lists of premises in later stages of the proof.

Let us first observe that if we do not care about the exact values of the polynomial bounds, an $O(\log t)$ height bound
along with a polynomial formula-size bound is all we need to show, as we will then get tree-like polynomial-size proofs
for free:
\begin{Lem}\label{lem:tree}
Let $\Pi$ be a dag-like $\fri$-derivation of $\fii$ from~$\Gamma$ of height~$h$ and formula size~$r$. Then there is a
tree-like $\fri$-derivation $\Pi'$ of $\fii$ from~$\Gamma$ of height~$h$ and formula size~$r$, hence with less than
$2^{h+1}$ lines and size $2^{h+1}r$. 
\end{Lem}
\begin{Pf}
We can unwind a dag-like derivation $\p{V,E,\gamma}$ with root~$\roo$ to a tree-like derivation $\p{V',E',\gamma'}$ of
the same height by taking for $V'$ the set of all paths ending in~$\roo$, with $\p{p,q}\in E'$ if $p$ initially
extends~$q$ by one edge, and $\gamma'_p=\gamma_v$ where $v$ is the starting vertex of~$p$. The bounds on the number of
lines and size follow from Observation~\ref{obs:param}.
\end{Pf}

Thus, a reader who is happy with any polynomial may ignore the exact bounds on the number of lines below and
concentrate on height bounds, which are easier to verify.

We need one more structural property of $\fri$-proofs so that we can accurately estimate the resulting proof size. Let
us say that an $\fri$-derivation is \emph{non-redundant} if no formula occurs in it more than once.
\begin{Lem}\label{lem:inf-size}
\
\begin{enumerate}
\item\label{item:3} Any $\fri$-derivation of $\fii$ from~$\Gamma$ can be made non-redundant by omitting some formulas.
\item\label{item:4} A non-redundant dag-like $\fri$-derivation of size~$s$ has inferential size $O(s)$.
\end{enumerate}
\end{Lem}
\begin{Pf}

\ref{item:3}: If we omit all but the first occurrence of each formula from a (sequence-like) $\fri$-derivation, it
remains an $\fri$-derivation.

\ref{item:4}: Clearly, the total size of axioms (logical or from~$\Gamma$) is at most~$s$. As for \eqref{eq:mp}
inferences, the size of an inference $\alpha,\alpha\to\beta\ru\beta$ is linear in the size of its second premise
$\alpha\to\beta$. In a non-redundant proof, each formula of the form $\alpha\to\beta$ can be used at most once as a
second premise of an \eqref{eq:mp} inference, because the conclusion of such an inference can only be~$\beta$, which
can only occur once in the derivation. Thus, the total size of \eqref{eq:mp} inferences is also $O(s)$.
\end{Pf}

We remark that property \ref{item:4} is specific to Frege systems based on \eqref{eq:mp} as the only rule of inference;
we see no reason it should hold in general. (Another such \eqref{eq:mp}-specific property is the last part of
Lemma~\ref{lem:sch}.)
\begin{Thm}\label{thm:fr-tree}
If $\fii$ has an $\fri$-derivation from~$\Gamma$ with $t$~lines, formula size~$r$, and size~$s$, then it has a
tree-like $\fri$-derivation from~$\Gamma$ with $O(t\log t)$ lines, height $O(\log t)$, formula size
$O(s+\LH\fii t)$, and size $O\bigl((s+\LH\fii t)(\log t)^2\bigr)$.
\end{Thm}
\begin{Pf}
Let $\Pi=\p{\gamma_i:i<t}$ be a derivation of~$\fii$ from~$\Gamma$, which we may assume to be non-redundant by
Lemma~\ref{lem:inf-size}. We fix $E\sset{<}\res[t]$ that makes $\p{[t],E,\gamma}$ a dag-like derivation by
Lemma~\ref{lem:f-seq-dag}. For each $j<t$ and $k\le\cl{\log t}$ such that $2^k\mid j$, we put
\begin{align*}
P_j^k&=\bigl\{i<j:\exists i'\in[j,j')\:\p{i,i'}\in E\bigr\},\\
\Gamma_j^k&=\p{\gamma_i:i\in[j,j')},\\
\Delta_j^k&=\p{\gamma_i:i\in P_j^k},\\
\tau_j^k&=\RET\fii{}\Delta_j^k\to\RET\fii{}\Gamma_j^k,
\end{align*}
where $j'=\min\{j+2^k,t\}$. Observe that $\lh{\Gamma_j^k}\le2^k$, $\lh{\Delta_j^k}=\lh{P_j^k}=O(2^k)$, and
$\lh{\tau_j^k}=O\bigl(\dlh{\Gamma_j^k}+\dlh{\Delta_j^k}+\LH\fii2^k\bigr)=O(s_{k,j}+\LH\fii2^k)$, where we put
$s_i=\lh{\gamma_i}+\sum_{\p{i',i}\in E}\lh{\gamma_{i'}}$ as in Definition~\ref{def:inf-size}, and
$s_{k,j}=\sum_{i\in[j,j')}s_i$. Notice that $\sum_is_i=O(s)$ by Lemma~\ref{lem:inf-size}, thus also $\sum_{2^k\mid
j}s_{k,j}=O(s)$ for each~$k$.

We construct $\fri^*$-derivations $\Pi_j^k$ of $\tau_j^k$ from~$\Gamma$ by induction on~$k$. For $k=0$, the formula
$\tau_j^0$ is $\RET\fii{}\p{\gamma_i:\p{i,j}\in E}\to\gamma_j^\fii$, which has a derivation from~$\Gamma$ with $O(1)$
lines and size $O(s_j+\lh\fii)$ using Lemma~\ref{lem:sch}. Assume that $\Pi_j^k$ have been defined for all $j<t$ such that
$2^k\mid j$, and let $j<t$ be such that $2^{k+1}\mid j$. If $j+2^k\ge t$, we have $\tau_j^{k+1}=\tau_j^k$, thus we can
take $\Pi_j^{k+1}=\Pi_j^k$. Otherwise, we combine $\Pi_j^k$ and $\Pi_{j+2^k}^k$ to $\Pi_j^{k+1}$ using an
$\fri^*$-proof of $\tau_j^k\to\tau_{j+2^k}^k\to\tau_j^{k+1}$ with $O(2^k)$ lines, height $O(\log t)$, formula size
$O(s_{k+1,j}+\LH\fii2^k)$, and size $O\bigl((s_{k+1,j}+\LH\fii2^k)\log t\bigr)$ that we construct as follows. Observe
that $\Gamma_j^{k+1}$ is the concatenation of $\Gamma_j^k$ and~$\Gamma_{j+2^k}^k$, and $\Delta_j^k\sset\Delta_j^{k+1}$,
while $\Delta_{j+2^k}^k$ is a concatenation of a subsequence of $\Delta_j^{k+1}$ and a subsequence of $\Gamma_j^k$.
Thus, Lemma~\ref{lem:subset-conj} gives us $\fri^*$-proofs of
\begin{align*}
\tau_j^k&\to\RET\fii{}\Delta_j^{k+1}\to\RET\fii{}\Gamma_j^k,\\
\tau_{j+2^k}^k&\to\RET\fii{}\Delta_j^{k+1}\to\RET\fii{}\Gamma_j^k\to\RET\fii{}\Gamma_{j+2^k}^k,\\
\RET\fii{}\Gamma_j^k\to\RET\fii{}\Gamma_{j+2^k}^k&\to\RET\fii{}\Gamma_j^{k+1}.
\end{align*}
with the stated size parameters. These together imply $\tau_j^k\to\tau_{j+2^k}^k\to\tau_j^{k+1}$.

In the end, $\Pi_0^{\cl{\log t}}$ is a derivation of $\top\to\RET\fii{i<t}\gamma_i$ from~$\Gamma$. This yields
$\RET\fii{}\p{\gamma_{t-1}}$, i.e., $\fii^\fii$, using Lemma~\ref{lem:subset-conj}, and we can infer~$\fii$.

It is clear that the whole derivation has height $O(\log t)$ and formula size $O(s+\LH\fii t)$. The derivations
$\Pi_j^0$ have together $O(t)$ lines and size $O\bigl(\sum_j(s_j+\lh\fii)\bigr)=O(s+\LH\fii t)$. Likewise, for each
$k<\cl{\log t}$, there are $t/2^{k+1}$ subproofs of $\tau_j^k\to\tau_{j+2^k}^k\to\tau_j^{k+1}$ with $O(2^k)$ lines
each, which together makes $O(t)$ lines of size
$O\bigl(\sum_{2^{k+1}\mid j}(s_{k+1,j}+\LH\fii2^k)\log t\bigr)=O\bigl((s+\LH\fii t)\log t\bigr)$. Summing over all
$k<\cl{\log t}$, the whole derivation has $O(t\log t)$ lines and size $O\bigl((s+\LH\fii t)(\log t)^2\bigr)$.
\end{Pf}
\begin{Rem}\label{rem:improve-fr}
We could avoid the machinery of $\RET\fii{}\Gamma$ formulas by defining
$\tau_j^k=(\Gamma_j^k\to\fii)\to(\Delta_j^k\to\fii)$, and using Lemmas \ref{lem:subset} and~\ref{lem:set-weak} in place of
Lemma~\ref{lem:subset-conj}, yielding an $\fri^*$-derivation with $O(t\log t)$ lines, height $O(\log t)$, formula size
$O(s)$, and size $O(st+\LH\fii t\log t)$.

If we have a real~$\land$, the $\lh\fii$ terms from the size parameters disappear: we obtain a derivation
with $O(t\log t)$ lines, height $O(\log t)$, formula size $O(s)$, and size%
\footnote{\cite[L.~4.4.8]{book} seemingly claims an even better bound $O(s\log t)$, but this is a typo, as the
argument only warrants size $O(st\log t)$; cf.\ \url{https://www.karlin.mff.cuni.cz/\string~krajicek/upravy.html}.}
$O\bigl(s(\log t)^2\bigr)$.

Back in the implicational setting, we can alternatively use $\RET p{}$ in place of $\RET\fii{}$, where $p$ is the
right-most variable occurrence in~$\fii$, i.e., $\fii$ is of the form $\Phi\to p$ for some sequence $\Phi$. This
reduces all the $\lh\fii$ terms in the size parameters to~$O(1)$: we obtain a derivation of $\fii^p$ from~$\Gamma$ with
$O(t\log t)$ lines, height $O(\log t)$, formula size $O(s)$, and size $O\bigl(s(\log t)^2\bigr)$. We can construct a
proof of $\fii^p\to\fii$ using Lemma~\ref{lem:set-weak}: two instances of~\eqref{eq:10} give
$\Phi\to(((\Phi\to p)\to p)\to p)\to p$, and \eqref{eq:15} yields $(((\Phi\to p)\to p)\to p)\to\Phi\to p$. We obtain an
$\fri^*$-derivation of $\fii$ from~$\Gamma$ with $O(t\log t+n)$ lines, height $O\bigl(\log(t+n)\bigr)$, formula size
$O(s)$, and size $O\bigl(s(\log t)^2+\LH\fii n\bigr)$, where $n=\lh\Phi\le\lh\fii$. Furthermore, if the $\IPC_\to$
tautology $\Gamma\to\fii$ is not a substitution instance of any strictly smaller $\IPC_\to$ tautology, then $n=O(t)$ because of
\cite[L.~4.4.4]{book}, which simplifies the bounds to $O(t\log t)$ lines, height $O(\log t)$, formula size $O(s)$, and
size $O\bigl(s(\log t)^2+\LH\fii n\bigr)$.

We can also modify the definition of $\RET\fii{}$ using an extra parameter~$\ell$ as in Remark~\ref{rem:improve-nm}. In the
context of the proof of Theorem~\ref{thm:fr-tree}, the optimal choice is $\ell\approx\sqrt{\LH\fii t(\log t)/s}$, which
yields an $\fri^*$-derivation of $\fii$ from~$\Gamma$ with $O(t\log t)$ lines, height $O(\log t)$, formula size
$O\bigl(s+\sqrt{\LH\fii st/\log t}\bigr)$, and size $O\bigl(s(\log t)^2+\sqrt{\LH\fii st(\log t)^3}\bigr)$.
\end{Rem}

Theorems \ref{thm:sim-nm-f} or~\ref{thm:sim-nm-f-conj}, \ref{thm:fr-tree}, and~\ref{thm:sim-f-nm} imply a polynomial
simulation of $\nmi$ by $\nmi^*$, but we can obtain better bounds by taking into account that the building blocks of
the proofs constructed in Theorems \ref{thm:sim-nm-f} and~\ref{thm:sim-nm-f-conj} are already tree-like:
\begin{Thm}\label{thm:nm-tree}
If $\fii$ has an $\nmi$-derivation from~$\Gamma$ with $t$~lines, size~$s$, and inferential size $\tilde s$, then it has
an $\fri^*$-derivation and an $\nmi^*$-derivation from~$\Gamma$ with $O(t^2)$ lines, height $O(\log t)$, formula size
$O(st)$, and size $O\bigl(\min\{st^2,\tilde st(\log t)^2\}\bigr)$.
\end{Thm}
\begin{Pf}
In view of Theorem~\ref{thm:sim-f-nm}, it suffices to construct an $\fri^*$-derivation.

We combine the arguments in Theorems \ref{thm:sim-nm-f-conj} and~\ref{thm:fr-tree}. Let $\Pi=\p{V,E,\gamma}$ be an $\nmi$-derivation of
$\fii$ from~$\Gamma$. By considering a topological ordering of $\p{V,E}$, we may assume $V=[t]$ and $E\sset{<}\res[t]$.
As in the proof of Theorem~\ref{thm:sim-nm-f-conj}, let $\p{\gamma'_i}_{i<t'}$, $t'\le t$, be an injective enumeration of
the set $\{\gamma_i:i<t\}$, and for each $i<t$, let $A'_i$ denote the sequence
$\p{\gamma'_j:j<t',\gamma'_j\in A_i\bez\Gamma}$. Put $\delta_i=\RET{\gamma_i}{}A'_i\to\gamma_i$; we have
$\lh{\delta_i}=O(s+\LH{\gamma_i}t)$.

Similarly to the proof of Theorem~\ref{thm:fr-tree}, for all $j<t$ and $k\le\cl{\log t}$ such that $2^k\mid j$, we put
\begin{align*}
P_j^k&=\bigl\{i<j:\exists i'\in[j,j')\:\p{i,i'}\in E\bigr\},\\
\Gamma_j^k&=\p{\delta_i:i\in[j,j')},\\
\Delta_j^k&=\p{\delta_i:i\in P_j^k},\\
\tau_j^k&=\RET\fii{}\Delta_j^k\to\RET\fii{}\Gamma_j^k,
\end{align*}
where $j'=\min\{j+2^k,t\}$. We have $\lh{\Gamma_j^k}\le2^k$ and $\lh{\Delta_j^k}=\lh{P_j^k}=O(2^k)$, thus
$\lh{\tau_j^k}=O\bigl(\dlh{\Gamma_j^k}+\dlh{\Delta_j^k}+\LH\fii2^k\bigr)=O(s2^k+s_{k,j}t)$, where
$s_{k,j}=\sum_{j\le i<j'}\lh{\gamma_i}+\sum_{i\in P_j^k}\lh{\gamma_i}\le s$. Observe $s_{k,j}\le\sum_{j\le
i<j'}s_{0,i}$, thus for a fixed~$k$, $\sum_js_{k,j}\le\sum_{i<t}s_{0,i}=\tilde s$. We will now construct
$\fri^*$-derivations $\Pi_j^k$ of $\tau_j^k$ from~$\Gamma$ by induction on~$k$.

As shown in the proof of Theorem~\ref{thm:sim-nm-f-conj}, for each $j<t$, there is an $\fri^*$-derivation of
$\Delta_j^0\to\delta_j$ from~$\Gamma$ with $O(t)$ lines, height $O(\log t)$, formula size $O(s+s_{0,j}t)$, and size
$O\bigl((s+s_{0,j}t)\log t\bigr)$. We can infer $\RET\fii{}\Delta_j^0\to\gamma_j^\fii$, which is $\tau_j^0$, using $O(1)$
extra lines of size $O(s+s_{0,j}t)$; we denote the resulting derivation $\Pi_j^0$. In total, these derivations have
$O(t^2)$ lines, height $O(\log t)$, formula size $O(rt)$ (where $r$ is the formula size of~$\Pi$) and size $O(\tilde
st\log t)$.

Let $k<\cl{\log t}$ and $j<t$ be such that $2^{k+1}\mid j$. If $j+2^k\ge t$, then $\tau_j^{k+1}=\tau_j^k$, and we put
$\Pi_j^{k+1}=\Pi_j^k$. Otherwise, we combine $\Pi_j^k$ and $\Pi_{j+2^k}^k$ to $\Pi_j^{k+1}$ using an $\fri^*$-proof of
$\tau_j^k\to\tau_{j+2^k}^k\to\tau_j^{k+1}$ as constructed in the proof of Theorem~\ref{thm:fr-tree}: it has $O(2^k)$ lines,
height $O(\log t)$, formula size $O(s2^k+s_{k+1,j}t)=O(st)$, and size $O\bigl((s2^k+s_{k+1,j}t)\log t\bigr)$; summing
this over all~$j$ for a fixed~$k$ gives $O(t)$ lines of total size $O(\tilde st\log t)$.

Altogether, $\Pi_0^{\cl{\log t}}$ has $O(t^2)$ lines, height $O(\log t)$, formula size $O(st)$, and size
$O\bigl(\tilde st(\log t)^2\bigr)$. It is a derivation of $\top\to\RET\fii{i<t}\delta_i$ from~$\Gamma$. Since
$\delta_{t-1}=\top\to\fii$, we can infer $\fii$ using Lemma~\ref{lem:subset-conj} without asymptotically increasing any of
the size parameters.

We can obtain the $O(st^2)$ size bound similarly, using $\delta_i=A'_i\to\gamma_i$ as in the proof of
Theorem~\ref{thm:sim-nm-f} in place of Theorem~\ref{thm:sim-nm-f-conj}; in this case, we can avoid usage of the $\RET\fii{}$
formulas entirely as in Remark~\ref{rem:improve-fr}.
\end{Pf}

We mention that if we have a real~$\land$, the size bound improves to $O\bigl(st(\log t)^2\bigr)$.

\end{document}